\documentclass[10pt,twocolumn,letterpaper]{article}

\usepackage[pagenumbers]{cvpr} 
\usepackage[accsupp]{axessibility}

\usepackage{enumitem}
\usepackage{multirow}
\usepackage{wrapfig}
\usepackage{colortbl}
\usepackage{dblfloatfix}
\usepackage[normalem]{ulem}
\usepackage{makecell}
\usepackage{url}
\usepackage{pifont}
\usepackage[hang,flushmargin]{footmisc}

\usepackage{algorithm}
\usepackage{algpseudocode}
\usepackage{microtype}

\usepackage[T1]{fontenc}
\usepackage[font=small,labelfont=bf,tableposition=top]{caption}
\DeclareCaptionLabelFormat{andtable}{#1~#2  \&  \tablename~\thetable}

\usepackage{pifont} 
\usepackage{xcolor} 
\usepackage{graphicx}

\newcommand{\cmark}{\textcolor{clova}{\ding{51}}} 
\newcommand{\xmark}{\textcolor{red}{\ding{55}}}   
\newcommand{\question}[1]{\noindent\textbf{#1}\,\,}

\setlist[itemize]{align=parleft,left=0pt,topsep=1mm,itemsep=0mm}

\definecolor{azure(colorwheel)}{rgb}{0.0, 0.5, 1.0}
\definecolor{nicegreen}{rgb}{0.0, 0.7, 0.1}
\definecolor{CuGray}{gray}{0.5}
\definecolor{rev}{rgb}{0.784, 0.003, 0.313}
\definecolor{pink}{cmyk}{0, 0.7808, 0.4429, 0.1412}
\definecolor{amethyst}{rgb}{0.6, 0.4, 0.8}
\definecolor{black}{rgb}{0.0, 0.0, 0.0}
\definecolor{tb3_yellow}{rgb}{0.996, 1.0, 0.6}
\definecolor{tb3_orange}{rgb}{0.980, 0.8, 0.604}
\definecolor{tb3_red}{rgb}{0.972, 0.6, 0.6}
\definecolor{blue}{rgb}{0.0, 0.0, 0.4}
\definecolor{clova}{rgb}{0.24, 0.63, 0.33}

\newcolumntype{g}{>{\columncolor{CuGray}}c}
\newcolumntype{z}{>{\columncolor{CuGray}}l}

\renewcommand{\paragraph}[1]{\vspace{1mm}\noindent\textbf{#1.}}

\usepackage{xspace}

\def\onedot{.\@\xspace}
\def\eg{\emph{e.g}\onedot} 
\def\ie{\emph{i.e}\onedot}

\def\etal{\emph{et al}\onedot}

\newcommand{\Sref}[1]{Sec.~\ref{#1}}
\newcommand{\Eref}[1]{Eq.~(\ref{#1})}
\newcommand{\Fref}[1]{Fig.~\ref{#1}}
\newcommand{\Tref}[1]{Table~\ref{#1}}

\newcommand{\calL}{{\mathcal{L}}}

\newcommand{\be}{\begin{eqnarray}}
\newcommand{\ee}{\end{eqnarray}}
\newcommand{\bee}{\begin{eqnarray*}}
\newcommand{\eee}{\end{eqnarray*}}

\newcommand{\matrixb}{\left[ \begin{array}}
\newcommand{\matrixe}{\end{array} \right]}

\newcommand{\para}[1]{\paragraph{#1}}

\definecolor{amethyst}{rgb}{0.6, 0.4, 0.8}

\usepackage{mathtools}

\definecolor{cvprblue}{rgb}{0.21,0.49,0.74}
\usepackage[pagebackref,breaklinks,colorlinks,allcolors=cvprblue]{hyperref}

\def\paperID{*****} 
\def\confName{CVPR}
\def\confYear{2025}

\title{Perceptually Accurate 3D Talking Head Generation: \\New Definitions, Speech-Mesh Representation, and Evaluation Metrics}

\def\authorBlock{
    Lee Chae-Yeon${}^{1*}$\enspace
    Oh Hyun-Bin${}^{2*}$\enspace
    Han EunGi${}^{1}$\enspace
    Kim Sung-Bin${}^{2}$\enspace
    Suekyeong Nam${}^{3}$\enspace
    Tae-Hyun Oh${}^{1,2,4}$\vspace{3mm} \\
   \small{${}^{1}$Grad. School of AI, POSTECH\quad${}^{2}$Dept. of Electrical Engineering,  POSTECH\quad${}^{3}$KRAFTON\quad${}^{4}$School of Computing, KAIST}\\ \vspace{-5mm}
}

\begin{document}
\author{\authorBlock}

\twocolumn[{%
\renewcommand\twocolumn[1][]{#1}%
\maketitle
\centering
    \centering
    \captionsetup{type=figure}
    \includegraphics[width=1.0\textwidth]{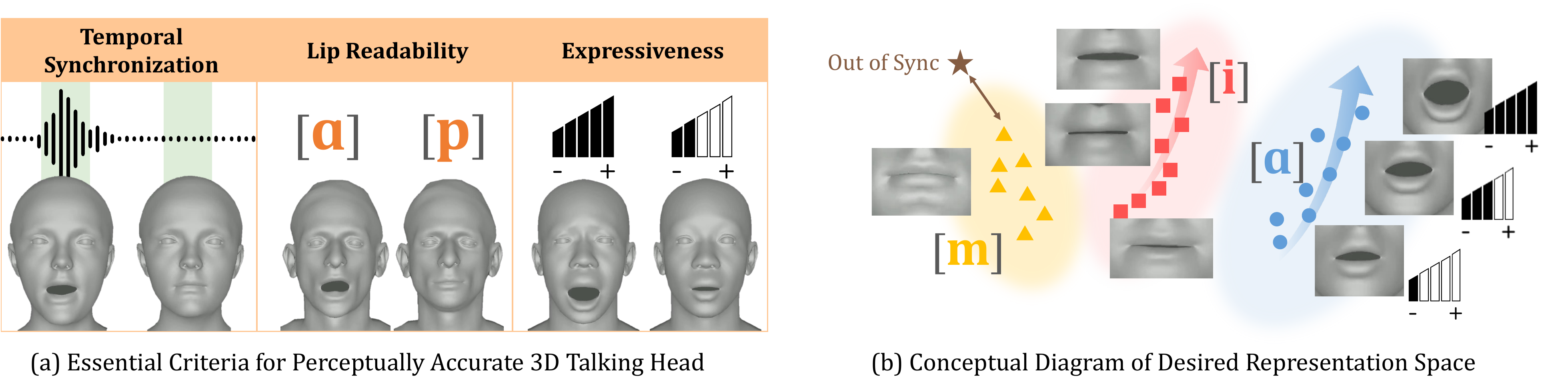}
    
    \vspace{-3mm}
    
    \caption{\textbf{What defines perceptually accurate lip movement for a speech signal?} In this work, we define three criteria to assess perceptual alignment between speech and lip movements of 3D talking heads: Temporal Synchronization, Lip Readability, and Expressiveness (a).
    The motivational hypothesis is the existence of a desirable representation space that models and complies well with the three criteria between diverse speech characteristics and 3D facial movements, as illustrated in (b); where
    representations with the same phonemes 
    are clustered, are sensitive to temporal synchronization, and follow a certain pattern
    as the speech intensity increases. Consequently, we build a rich speech-mesh synchronized representation space that exhibits the desirable properties.
    }
    \vspace{6mm}
    \label{fig:teaser}
}]

\def\thefootnote{*}\footnotetext{These authors contributed equally.}
\begin{abstract}
Recent advancements in speech-driven 3D talking head generation have made significant progress in lip synchronization. However, existing models still struggle to capture the perceptual alignment between varying speech characteristics and corresponding lip movements. In this work, we claim that three criteria—Temporal Synchronization, Lip Readability, and Expressiveness—are crucial for achieving perceptually accurate lip movements. 
Motivated by our hypothesis that a desirable representation space exists to meet these three criteria, 
we introduce a speech-mesh synchronized representation that captures intricate correspondences between speech signals and 3D face meshes. 
We found that our learned representation exhibits desirable characteristics, and 
we plug it into existing models
as a perceptual loss to better align lip movements to the given speech.
In addition, we utilize this representation as a perceptual metric and introduce two other physically grounded lip synchronization metrics to assess how well the generated 3D talking heads align with these three criteria. 
Experiments show that training 3D talking head generation models with our perceptual loss significantly improve all three aspects of perceptually accurate lip synchronization. Codes and datasets are available at \href{https://perceptual-3d-talking-head.github.io/}{https://perceptual-3d-talking-head.github.io/}.
\end{abstract}
\vspace{-2mm}    
\section{Introduction}
\label{sec:introduction}
Speech-driven 3D talking head generation focuses on generating 3D facial movements synchronized with input speech signals. 
This plays a key role in enhancing communication within multimedia applications, such as virtual reality, entertainment, and education~\cite{acceptability}. 
To provide users with a more realistic and immersive experience, it is crucial that the facial and lip movements of 3D avatars are synchronized with the various aspects of speech. This synchronization should be 
perceptually
accurate from a human perspective, ensuring that the avatars' expressions are both natural and convincing.

While 
recent works in learning-based 3D talking head generation~\cite{voca,meshtalk,codetalker,selftalk,facediffuser,emote} aim to 
enhance the
lip synchronization capabilities,
they commonly rely on minimizing the Mean Squared Error (MSE) loss between generated 3D facial 
motion and 
ground truth motion
as a learning objective. This approach is practical, as it directly contributes to minimizing the Lip Vertex Error (LVE), a commonly used metric that measures the MSE between the lip vertices of generated 3D facial motions and the ground truth.

Despite the improvements in the LVE metric, existing models still struggle to correlate lip movements with some
speech characteristics, such as wider mouth openings as speech volume increases. These characteristics are not adequately captured by existing datasets~\cite{voca, biwi}, which have limited ranges of facial motion patterns due to the small dataset scale and restricted intensity range. 
Moreover, relying on MSE and LVE is insufficient for learning or assessing perceptually plausible lip motions~\cite{selftalk, sung2024multitalk}, as it focuses solely on vertex-wise geometric differences and overlooks the true correspondence between the speech signals and lip movements.
A lower MSE and LVE do not necessarily correspond to a more perceptually accurate lip movement.

These observations raise critical questions: \emph{What defines perceptually accurate lip movement in response to a speech signal, and how can we enhance this accuracy?}
We draw inspiration from findings on human audio-visual perception: 1) Humans are sensitive to temporal asynchrony between speech and lip movements; slight discrepancies can disrupt the perception of natural synchronization~\cite{vatakis2006audiovisual}. 2) Humans rely on accurate viseme-phoneme correspondence when assessing lip-sync accuracy, expecting visual lip movements to match the spoken phonemes~\cite{bear2017phoneme}. 3) There is a proportional increase in jaw and lip movements as speech intensity increases, contributing to the expressiveness perceived in natural speech~\cite{suma2023effects,huber2006effects,tasko2004variations,schulman1989articulatory}.
Through our human study, we reveal an intriguing finding: participants favor lip movements with intensity that corresponds to speech---even if they exceed the established maximum acceptable asynchrony~\cite{vatakis2006audiovisual} by twice---over those that are perfectly synchronized but lack expressive alignment (\Tref{table:human_study}-[Right]). This reveals that humans are more sensitive to expressiveness than temporal synchronization when perceiving, highlighting the importance of expressiveness.
Building upon these insights, we define three criteria that significantly impact the perceptual lip synchronization of 3D talking heads: \emph{Temporal Synchronization}, \emph{Lip Readability}, and \emph{Expressiveness} (\Fref{fig:teaser} (a)).

Motivated by our hypothesis that a desirable representation space exists to meet these three criteria,
we propose a speech-mesh synchronized and rich representation that captures the intricate correspondence between speech and 3D face mesh. 
We design a transformer-based architecture that maps time-sequenced speech and mesh inputs to a shared representation space. To effectively train this system, we employ a two-stage training method: we first develop a robust audio-visual speech representation using a large-scale 2D video dataset~\cite{lrs3}, which then serves as an anchor for learning a speech-mesh representation. We found that the first step is important and leads to emergent desirable properties 
(illustrated in Fig.~\ref{fig:teaser} (b))
in the final representation.
This sequential approach ensures that the model first establishes a space that captures a wide range of speech characteristics, and then extends to explore the relationships between speech and 3D face mesh across diverse speech intensities and facial movements.
Adopting this representation, we introduce a plug-and-play perceptual loss adaptable to any existing 3D talking head generation models ~\cite{faceformer, codetalker, selftalk}, enhancing the perceptual quality of 3D talking heads.

Furthermore, to assess the three criteria, we introduce three metrics for each aspect.
We leverage our learned representation as a perceptual metric, Perceptual Lip Readability Score (PLRS), to evaluate the perceptual lip readability of lip movements. 
Also, we propose two physically grounded lip synchronization metrics: Mean Temporal Misalignment (MTM) for temporal synchronization and Speech-Lip Intensity Correlation Coefficient (SLCC) for expressiveness.

Extensive experiments demonstrate 
that our perceptual loss significantly enhances all three aspects: Temporal Synchronization, Lip Readability, and Expressiveness, which are demonstrated across various metrics: existing 
metric, our newly proposed metrics, and human evaluations. 
We also find that incorporating an additional pseudo-dataset~\cite{mead}, which captures diverse ranges of speech and lip movement intensities, can further improve expressiveness.
Our main contributions are summarized as follows:
\begin{itemize}
    \item Defining three aspects—Temporal Synchronization, Lip Readability, and Expressiveness—that affect the perceptual quality of 3D talking heads and proposing three evaluation metrics for these aspects.
    \item Constructing a speech-mesh representation space that captures rich and diverse correspondences between speech and lip movements.
    \item Proposing a plug-in perceptual loss using the constructed speech-mesh representation and demonstrating improvements on existing metric, our newly proposed metrics, and human evaluations.
\end{itemize}

\section{Related Work}\vspace{-1mm}
\label{sec:related_work}


\paragraph{Speech-driven 3D talking head generation}
Speech-driven 3D talking head generation aims to generate realistic 3D facial movements aligned with given speech. 
Among recent data-driven methods~\cite{voca, faceformer, codetalker, selftalk, emotalk, emote, yang2024probabilistic, facediffuser},
FaceFormer~\cite{faceformer} introduces a transformer-based autoregressive model and leverages a pre-trained speech model to capture long-term audio context and 
past facial movements. CodeTalker~\cite{codetalker} employs a VQ-VAE 
to construct a discrete facial motion space, addressing the over-smoothing problem. Diffusion models~\cite{ddpm,ddim} have also been demonstrated to be effective for 3D talking head generation~\cite{facediffuser,diffposetalk}. In addition to synthesizing neutral facial motions, several works extend the 3D talking head to express specific aspects, such as emotional expressions~\cite{emote,emotalk}, multilingual capabilities~\cite{sung2024multitalk}, or laughter~\cite{laughtalk}.
Despite these advances, existing methods rely on minimizing MSE loss without a clear definition of perceptually accurate lip movement, overlooking the multifaceted nature of lip synchronization.
To address this, we define three critical aspects of lip synchronization
and propose rich speech-mesh representation, along with its 
application as a perceptual loss in a plug-and-play manner, enhancing all three aspects of lip synchronization quality in existing 3D talking head generation models~\cite{faceformer, codetalker, selftalk}.

\paragraph{Speech-face representation learning}
Well-aligned representation spaces learned from large-scale datasets such as CLIP~\cite{clip} and ImageBind~\cite{imagebind} are valued for their scalability and versatility.
These spaces enable a wide range of applications, including auxiliary loss~\cite{tewel2022zerocap}, intermediate representation~\cite{mdm}, and evaluation metrics~\cite{hessel2021clipscore}. 
With this context, audio-visual representation spaces trained specifically on speech and 2D face videos have been proposed~\cite{syncnet, avhubert, cavmae, hicmae}. 
For instance, SyncNet~\cite{syncnet}, a CNN-based model learns to detect audio-visual temporal synchronization and has been applied to several tasks, such as active speaker detection~\cite{lrs3,chung2018voxceleb2} and 2D talking head generation~\cite{wav2lip,wang2023seeing,sadtalker}. 
Similarly, a transformer-based AV-HuBERT~\cite{avhubert} has demonstrated remarkable effectiveness in various tasks, including lip reading~\cite{shi2022robust}, audio-visual translation~\cite{choi2024av2av}, and 2D talking head generation~\cite{wang2023seeing}.
While there has been significant progress in the 2D domain, advancements in the 3D domain remain under-explored.
Yang~\etal~\cite{yang2024probabilistic} extends the SyncNet architecture to accommodate speech and 3D face meshes; however, its application is limited to evaluating 3D talking heads. 
In this work, we demonstrate the versatility of our representation space as a plug-in module to improve the perceptual accuracy of the existing speech-driven 3D talking head generation models~\cite{faceformer, codetalker, selftalk}, as well as to assess their performance. 

\paragraph{Evaluation metrics for speech-driven 3D talking head}
The prevalent evaluation metric, Lip Vertex Error (LVE)~\cite{meshtalk}, measures the L2 distance between predicted lip vertices and ground truth. 
Additional metrics, such as Upper Face Dynamics Deviation (FDD)~\cite{codetalker} and Lip Readability Percentage (LRP)~\cite{selftalk}, consider different regions of facial motion.
These metrics, however, focus on vertex-wise geometric differences between the ground-truth 3D facial motions and neglect speech-related information.
To incorporate both speech and 3D face mesh data for evaluation, MultiTalk~\cite{sung2024multitalk} introduces an Audio-Visual Lip Readability (AVLR) metric, which assesses perceptual accuracy of lip readability using a pre-trained Audio-Visual Speech Recognition model~\cite{muavic}. Yet, AVLR relies on speech and 2D face video rendered from the 3D face mesh, which may not align with the 3D talking head domain.
To address these limitations, we introduce novel and comprehensive evaluation metrics that focus on diverse aspects of lip synchronization: Mean Temporal Misalignment (MTM), Perceptual Lip Readability Score (PLRS), and Speech-Lip Intensity Correlation Coefficient (SLCC).
\section{Essential Criteria for Perceptually Accurate 3D Talking Head}
\label{sec:definition}
Generating a 3D talking head with lip movements that are perceptually accurate to human observers requires a clear understanding of the components that influence perceptual quality.
Although existing works have focused on improving partial aspects of these talking head generation models, a comprehensive exploration 
for establishing a perceptually accurate 3D talking head model has barely been undertaken. Drawing inspirations from extensive research in existing works~\cite{vatakis2006audiovisual,wang2023seeing, selftalk, eungi24_interspeech,suma2023effects, huber2006effects, tasko2004variations, schulman1989articulatory} and our 
studies, we identify three fundamental criteria essential for achieving perceptually accurate lip movements in 3D talking heads.

\paragraph{Temporal Synchronization}
This alignment is particularly crucial in media involving human speech, where viewers expect lip movements to precisely match the corresponding speech in time.
Temporal misalignment between speech and lip movements indeed distracts viewers, reduces user experiences, and negatively impacts audience perception~\cite{reeves1993effect}. 
Vatakis~\etal\cite{vatakis2006audiovisual} find that viewers are particularly sensitive to speech-lip asynchrony compared to other audio-visual asynchrony, such as music. They note that mismatches become noticeable when speech precedes lip movements by more than 50 ms or follows them by more than 220 ms.
These findings may hold the same in 3D talking head generation, where any misalignment between 
speech and lip movements
can break the illusion of realism, leading to a diminished user experience and reducing the perceived authenticity of the virtual character.
Thus, we define the temporal synchronization of the talking head as an important aspect of lip synchronization quality.

\newcommand\JUGE{\fontsize{100}{120}\selectfont}
\newcommand\HUGE{\fontsize{56}{70}\selectfont}

\paragraph{Lip Readability}
Visemes (or lip movements) must correspond accurately to the speech phonemes to ensure the spoken words are visually intelligible~\cite{bear2017phoneme}.
This aspect is widely acknowledged as important in existing literature in both 2D~\cite{wang2023seeing} and 3D~\cite{selftalk, eungi24_interspeech} speech-driven talking head models, which leverage lip reading experts as an auxiliary guidance to improve the visual intelligibility of the spoken word.
However, the mapping between speech and lip movements is not one-to-one, making lip readability challenging to define. 
For example, the size and shape of the opening mouth
and the dynamic movement patterns of the lips in response
to specific utterance differ at every moment~\cite{yang2024probabilistic} or per individual~\cite{song2024talkingstyle}.
To capture this complexity, we define the lip readability within speech-mesh synchronized representation space learned from a large-scale dataset, capturing the comprehensive and nuanced correspondence between speech and lip movements.

\begin{table}[t]
    \centering
    \begin{minipage}{.5\linewidth}
        \centering
        \includegraphics[width=1.0\linewidth]{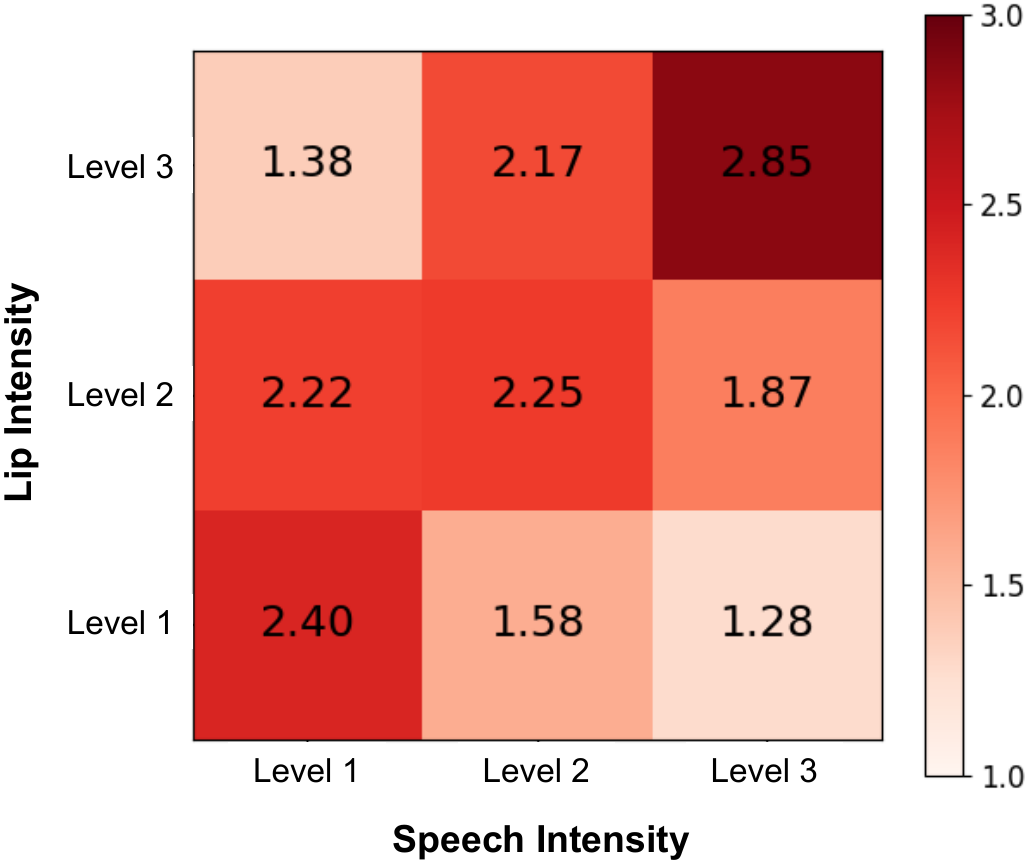}
    \end{minipage}
    \hfill
    \begin{minipage}{.4\linewidth}
        \centering
        \renewcommand{\arraystretch}{1.05} 
        \resizebox{.97\linewidth}{!}{
            \Huge
            \begin{tabular}{c c c}
            \toprule
            & \multicolumn{2}{c}{Samples} \\
            \cmidrule(lr){2-3}
            & A & B \\
            \midrule
             Temp. & \cmark & \xmark \\
             Exp. & \xmark & \cmark \\
             \midrule
             Prefer (\%) & 17.4 & 82.6 \\
            \bottomrule
            \end{tabular}}
    \end{minipage}
    \vspace{1mm}
    \caption{\textbf{Human studies on alignment criteria.} [Left] Preference scores (1-3) for 3D talking heads with varying lip movement intensities paired with different speech intensities. [Right] Human preference between (A) samples with precise timing but low expressiveness, and (B) samples with high expressiveness but 100ms asynchrony—twice the commonly accepted 50ms threshold~\cite{vatakis2006audiovisual}.}
    \label{table:human_study}
\vspace{-4mm}
\end{table}

\paragraph{Expressiveness}
Speech conveys not only linguistic content but also varies in intensity. For instance, a speaker may express the same text softly or loudly, with jaw and lip movements proportionally increasing as speech intensity rises, which contributes to perceived expressiveness in real-world face recognition~\cite{huber2006effects,tasko2004variations,schulman1989articulatory}.
To demonstrate that the positive 
correlation of human preference between the intensity of speech and lip movements also exists in 3D talking head field, we conduct a human study by presenting 3D talking heads with varied lip movement intensities paired with different speech intensities in 
\Tref{table:human_study}-[Left].
Participants prefer the lip movements with the intensity that match the intensity of speech.
Despite this distinct human preference for synchronization between the intensity of corresponding speech and lip movements, this aspect has barely been explored in the talking head generation field.
We thus define the expressiveness in 3D talking head as the correlation between speech and lip movement intensity, which is crucial for establishing genuine talking heads and is expected to guide future research aimed at improving perceptual lip synchronization.

\paragraph{Focus of this work}
Among the three criteria for perceptually accurate 3D talking head, we find that recent 3D talking head generation models achieve reasonable temporal alignment between speech and lip movements
(see supplementary for a visualization of temporal synchronization).
Furthermore, we design an A vs. B test, prompting participants to choose between two samples: (A) temporally synchronized one while
lacking expressive synchronization, and (B) the other with expressive synchronization but with 100ms asynchrony.
\Tref{table:human_study}-[Right] shows that users prefer sample B, suggesting that humans may prioritize expressive synchronization over strict temporal alignment.
This insight directs our focus toward enhancing all three aspects
to better capture perceptual realism in 3D talking head.
Details of the human study are in the supplementary material.
Before introducing metrics to assess the criteria, we present our synchronized representation for designing our perceptual loss.

\begin{figure*}[t]
    \centering
    \vspace{-2mm}
    \includegraphics[width=0.875\linewidth]{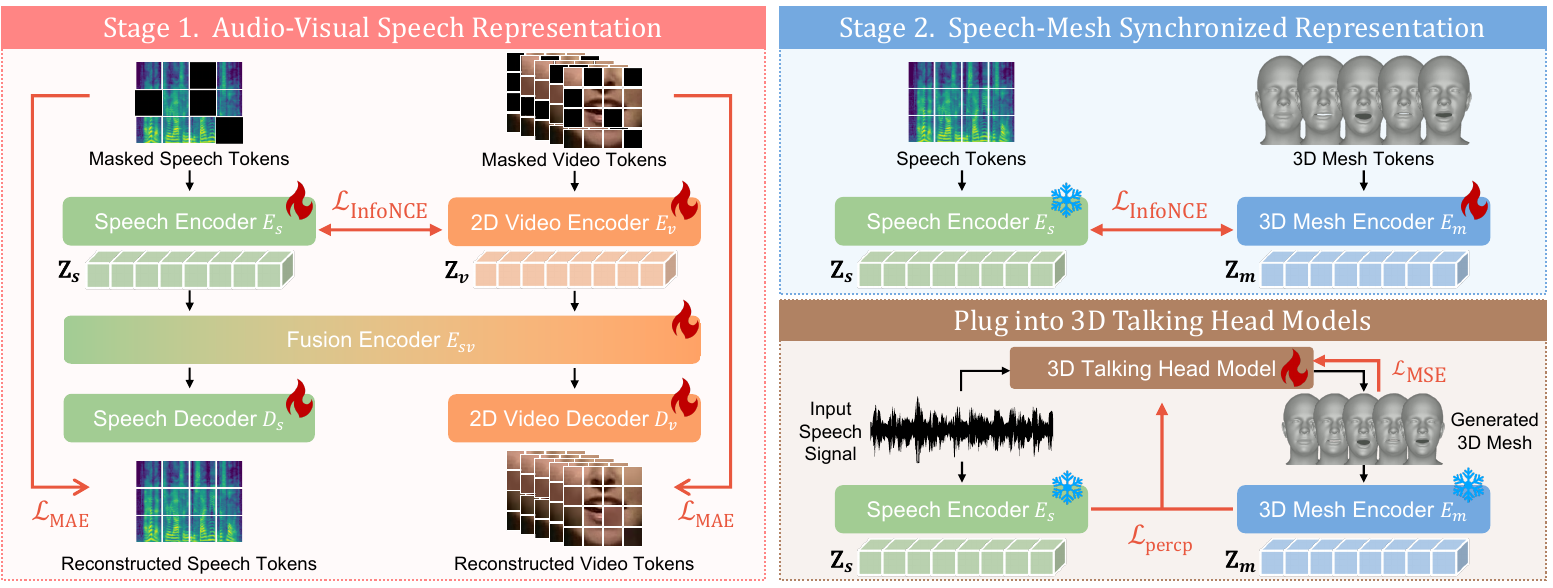}\vspace{-2.5mm}
    \caption{\textbf{Pipeline of speech-mesh synchronized representation learning.} We train our speech-mesh representation space in a two-stage manner. In the first stage, we learn a rich audio-visual 
    representation in 2D domain to capture the synchronization between lip movement and speech.
    In the second stage, we train the 3D mesh encoder to align the 3D mesh space with the frozen speech space. As an application of our speech-mesh representation space, we propose a plug-in perceptual loss to 3D talking head models to enhance the quality of lip movements.
    }
    \label{fig:pipeline}
    \vspace{-4mm}
\end{figure*}

\section{Speech-Mesh Synchronized Representation}
\label{sec:method}

We hypothesize that a desirable representation space exists to meet the three criteria defined in \Sref{sec:definition}. Motivated by this hypothesis, we develop a rich speech-mesh synchronized representation which captures the intricate correspondence between speech and the 3D face mesh. We found that our learned representation exhibits desirable properties, and we adopt it as a perceptual loss to improve the perceptual accuracy of existing models with respect to these three criteria.

\para{Overview}
Directly learning a synchronized speech-mesh representation presents challenges due to the scarcity of speech-3D face mesh datasets. One potential solution is to construct pseudo-GT 3D face meshes using reconstruction models~\cite{spectre}, although relying solely on pseudo-GT may not suffice for building a robust representation. To overcome this, we leverage the extensive knowledge from the speech-2D lip video representations and adapt this to accommodate 3D face meshes. Specifically, we propose a two-stage training process: in stage 1,
we learn an audio-visual speech representation that accurately reflects lip movements from the unlabeled in-the-wild 2D synchronized talking face video dataset, LRS3~\cite{lrs3}. Subsequently, we leverage the pre-trained speech representation from stage 1,
using it as the anchor space to learn a synchronized speech-mesh representation.

\para{Stage 1. Learning audio-visual speech representation}
This stage aims to learn a rich speech-2D lip video representation that effectively captures the correlation between varying speech characteristics and lip dynamics. Motivated by prior works~\cite{cavmae, hicmae}, we extend 
the integration of masked autoencoder (MAE) and cross-modal contrastive learning to learn a synchronized speech-2D lip video representation using 2D videos. The architecture for stage 1 consists of two modality-specific encoders, a cross-modal fusion encoder, and two modality-specific decoders.
(see \Fref{fig:pipeline}-[Stage 1]).

Given a speech and 2D lip video pair, $(\mathbf{X_s}, \mathbf{X_v}) \in D$, we begin by patchifying and tokenizing speech spectrograms and video frames into speech and video tokens as $\mathbf{S} = (\mathbf{s}_1, \dots, \mathbf{s}_N)$ and $\mathbf{V} = (\mathbf{v}_1, \dots, \mathbf{v}_M)$, where $\mathbf{s}_{i}, \mathbf{v}_{j} \in \mathbb{R}^{H}$.
We then randomly mask out $P$\% portion of tokens. The remaining unmasked speech tokens $\mathbf{S}^{unmask}$ and video tokens $\mathbf{V}^{unmask}$ are respectively fed into the speech encoder ${E}_s$ and video encoder ${E}_v$, each consisting of $N_e$ transformer layers.
Each encoder extracts uni-modal embeddings, $\mathbf{Z}^l_\mathbf{s}={E}^l_s(\mathbf{S}^{unmask})$ and $\mathbf {Z}^l_\mathbf{v}={E}^l_v(\mathbf{V}^{unmask})$, where ${l} \in (1, \dots, N_e)$ denote the layer indices. Also, the mean pooled speech and video embeddings, $\mathbf{c}^l_\mathbf{s}=\text{MeanPool}(\mathbf{Z}^l_\mathbf{s})$ and $\mathbf{c}^l_\mathbf{v}=\text{MeanPool}(\mathbf{Z}^l_\mathbf{v})$, are derived from the corresponding uni-modal embeddings.
Following the uni-modal encoders, we introduce a multi-modal fusion encoder ${E}_{sv}$ to exploit complementary information from each extracted uni-modal embedding. The speech and video embeddings are jointly processed through this encoder, resulting in multi-modal fusion embeddings $\mathbf{F_s}$ and $\mathbf{F_v}$, \ie, $[\mathbf{F_s},\mathbf{F_v}]={E}_{sv}([\mathbf{Z_s},\mathbf{Z_v}])$, where $[,]$ denotes concatenation.
With the fusion embeddings $\mathbf{F_s}$ and $\mathbf{F_v}$, we utilize each modality's decoder to reconstruct the original signals. Specifically, we employ a speech decoder ${D}_s$ and a video decoder ${D}_v$, each consisting of $N_d$ transformer layers. We pad each multi-modal fusion embedding with trainable masked tokens at their original positions, resulting in $\mathbf{F_s'}$ and $\mathbf{F_v'}$. Each decoder then reconstructs the respective signals; the reconstructed speech spectrogram and video tokens as $\hat{\mathbf{S}}^{mask}{=}{D}_s(\mathbf{F_s'})$ and $\hat{\mathbf{V}}^{mask}{=}{D}_v(\mathbf{F_v'})$.

Training the model involves two objectives: learning the cross-modal alignment between the speech-2D lip video signals, while reconstructing their original signals.
Given $B$ speech-2D lip video token pairs in a batch, $\left\{\mathbf{S}_i, \mathbf{V}_i\right\}_{i=1}^{B}$, we treat the true speech-2D lip video pair as positive samples, while the others in the batch are considered negative samples.
To encourage alignment between temporally synchronized speech-2D lip video instances, we utilize a cross-modal contrastive learning strategy, InfoNCE~\cite{oord2018representation}. 
We maximize the cosine similarity between positive sample of the mean pooled speech embedding $\mathbf{c}^l_{\mathbf{s},i}=\text{MeanPool}({E}^l_s(\mathbf{S}_{i}^{unmask}))$ and the corresponding synchronized mean pooled video embedding $\mathbf{c}^l_{\mathbf{v},i}=\text{MeanPool}({E}^l_v(\mathbf{V}_{i}^{unmask}))$. 
We first define the speech-centric loss as:
\begin{equation}
\label{eq:atovloss}
\calL_{\text{S} \rightarrow \text{V}}=-\frac{1}{B}\sum_{i=1}^{B}\log\tfrac{\exp(\mathbf{c}_{\mathbf{s},i}^{l}\cdot \mathbf{c}_{\mathbf{v},i}^{l}/\tau)}{\sum_{j=1}^{B}\exp(\mathbf{c}_{\mathbf{s},i}^{l}\cdot \mathbf{c}_{\mathbf{v},j}^{l}/\tau)},
\end{equation}
where $\tau$ is a temperature hyperparameter. 
Also, 
we make the objective symmetric by defining a video-centric loss as $\calL_{\text{V} \rightarrow \text{S}}$. 
We sum $\calL_{\text{S} \rightarrow \text{V}}$ and $\calL_{\text{V} \rightarrow \text{S}}$ across $L$ selected encoder layers to obtain our final contrastive learning objective:
%
\begin{align}
\label{eq:infonceloss}
\calL_{\text{InfoNCE}}=\sum\nolimits_{l \in L}\calL_{\text{S} \rightarrow \text{V}}+\calL_{\text{V} \rightarrow \text{S}}.
\end{align}
For reconstruction, the model is trained in a self-supervised manner
by minimizing the reconstruction loss
$\calL_{\text{MAE}}$ 
as:
\begin{equation}
\label{eq:maeloss}
\resizebox{\linewidth}{!}{%
$\calL_{\text{MAE}}\!=\!
\frac{1}{B}\sum_{i=1}^{B}\!\left[\!\frac{\sum
\|\hat{\mathbf{S}}_{i}^{mask}-\mathbf{S}_{i}^{mask}\|_F^2}{|\mathbf{S}_{i}^{mask}|}
\!+\!
\frac{\sum
\|\hat{\mathbf{V}}_{i}^{mask}-\mathbf{V}_{i}^{mask}\|_F^2}{|\mathbf{V}_{i}^{mask}|}\!\right]$},
\end{equation}
where $|\mathbf{S}_{i}^{mask}|$ and $|\mathbf{V}_{i}^{mask}|$ denote the number of masked speech and video tokens, respectively.
Overall, our 
objective function 
is defined as:
\vspace{-1mm}
\begin{equation}
\label{eq:totalloss}
\calL=\calL_{\text{MAE}}+\lambda \calL_{\text{InfoNCE}},
\vspace{-1mm}
\end{equation}
where $\lambda$ is the weight factor for cross-modal contrastive loss.

We observe that the representation space trained with
the transformer architecture and a rich and large-scale 2D face video dataset in this way already possesses the desired properties we pursue, illustrated in \Fref{fig:teaser} (see supplementary for visualizations of pre-trained speech representation). 
Motivated by this, we transfer these emergent properties to the speech-mesh representation space as follows.

\para{Stage 2. Learning speech-mesh representation}
In this stage, we design a 3D mesh encoder that maps 3D face mesh to the speech representation pre-trained in stage 1. This pre-trained speech representation, derived from rich speech-2D face video data, serves as a robust anchor space for learning the correlation between diverse speech characteristics and 3D facial motions. We use contrastive learning loss to align 3D facial motion embeddings with the anchored speech representations, as shown in \Fref{fig:pipeline}-[Stage 2].

Given a speech and 3D face mesh pair $(\mathbf{X_s}, \mathbf{X_m}) \in D$, we patchify and tokenize speech spectrograms into speech tokens $\mathbf{S}$, and map these into pre-trained uni-modal mean pooled speech embeddings $\mathbf{c}^l_\mathbf{s}=\text{MeanPool}({E}^l_s(\mathbf{S}))$. 
Similarly, we patchify 3D face mesh into mesh tokens $\mathbf{M}$, and feed them into the 3D mesh encoder ${E}_{m}$ consisting of $N_e$ transformer layers to extract mesh embeddings $\mathbf{Z}^l_\mathbf{m}={E}^l_m(\mathbf{M})$ and the corresponding mean-pooled mesh embedding $\mathbf{c}^l_\mathbf{m}$. 
For the learning objective, given $B$ speech-3D face mesh token pairs in a batch, $\left\{\mathbf{S}_i, \mathbf{M}_i\right\}_{i=1}^{B}$, we first define speech-anchored loss as:
\begin{equation}
\label{eq:atomloss}
\calL_{\text{S} \rightarrow \text{M}}=-\frac{1}{B}\sum_{i=1}^{B}\log\tfrac{\exp(\mathbf{c}_{\mathbf{s},i}^{l}\cdot \mathbf{c}_{\mathbf{m},i}^{l}/\tau)}{\sum_{j=1}^{B}\exp(\mathbf{c}_{\mathbf{s},i}^{l}\cdot \mathbf{c}_{\mathbf{m},j}^{l}/\tau)},
\end{equation}
as well as $\calL_{\text{M} \rightarrow \text{S}}$ which is the mesh-anchored loss. Similar to \Eref{eq:infonceloss}, we sum $\calL_{\text{S} \rightarrow \text{M}}$ and $\calL_{\text{M} \rightarrow \text{S}}$ across $L$ selected encoder layers to train the 3D mesh encoder ${E}_{m}$ while the speech encoder fixed. 
As a result, this two-stage training approach yields a robust speech-mesh representation, which significantly enhances the alignment between the speech and 3D face mesh modalities.

\para{Adopting it as a perceptual loss}
A key application of the speech-mesh representation learned through the two-stage training process is its use as a perceptual loss to enhance the perceptual accuracy of the 3D talking head model (see \Fref{fig:pipeline}). Leveraging this representation as a perceptual loss ensures that the generated lip movements are perceptually accurate and aligned well 
with the speech.

Given $B$ speech and generated 3D face mesh pairs in a batch, $\{\mathbf{S}_i, \hat{\mathbf{M}}_i\}_{i=1}^{B}$, we define our perceptual loss with the symmetric InfoNCE loss 
(\Eref{eq:infonceloss}) as
\begin{align}
    \label{eq:percploss}
    \calL_{percp}=\sum_{l \in L} \calL_{\text{S} \rightarrow \text{M}}+\calL_{\text{M} \rightarrow \text{S}}.
\end{align}
Our perceptual loss encourages synchronized speech and mesh embeddings to pull closer together, while unsynchronized ones to push apart.
The effectiveness and analysis of our perceptual loss will be discussed in later sections.
\section{Evaluation Metrics}
\label{sec:evalution}

In this section, we describe our proposed evaluation metrics that assess each criterion impacting the quality of 3D lip accuracy, as discussed in \Sref{sec:definition}. Here, we outline the high-level concepts of our proposed metrics. Additional details and experiments are provided in the supplementary material.

\paragraph{Mean Temporal Misalignment (MTM)}
To measure the temporal discrepancy between speech and corresponding lip movements, temporal correspondence annotations, such as onset times for each modality, would typically be required. 
As a proxy, we determine temporal correspondence and measure temporal discrepancies between the ground truth and predicted lip vertex displacement sequences by using Derivative Dynamic Time Warping (DDTW)~\cite{keogh2001derivative}, which robustly identifies local structural similarities compared to standard Dynamic Time Warping (DTW)~\cite{berndt1994using}.
For simplicity, we focus on the central vertices of the upper and lower lips when extracting displacement sequences, and use local extrema in the DDTW process to measure temporal misalignment by pinpointing precise time steps for mouth opening and closing events.
Consequently, Mean Temporal Misalignment (MTM) $\overline{\Delta t}$ is defined as 
$\overline{\Delta t} = \tfrac{1}{K} \sum\nolimits_{k=1}^{K} \Delta t_k$,
where $K$ is the total number of video clips and $\Delta t_k$ is the averaged temporal misalignment of the $k$-th video clip.

\paragraph{Perceptual Lip Readability Score (PLRS)}
While Lip Vertex Error (LVE) measures the accuracy of generated lip articulations against the ground truth, it does not fully assess whether lip movements are perceptually aligned with the given speech. To address this, we leverage our speech-mesh representation in Sec.~\ref{sec:method} as a perceptual lip readability evaluator. We compute the perceptual alignment using the cosine similarity of the mean pooled speech and mesh embeddings. Since this representation has learned a rich distribution of speech correspondences across various facial movements, our metric correlates highly with human perception, measuring perceptual lip movement alignment more accurately than LVE (see supplementary for the human study on metrics).

\paragraph{Speech and Lip Intensity Correlation Coefficient (SLCC)}
As shown in \Tref{table:human_study}-[Left],
humans prefer aligned intensity between speech and lip movement.
Thus, the intensity of generated lip movements should positively correlate with the corresponding input speech.
To quantify this, we define the Speech and Lip Correlation Coefficient $r_{SL}$ as:
\begin{equation}
\label{eq:LI}
r_{SL} = \tfrac{\sum_{k=1}^{K} (\text{SI}_{k} - \bar{\text{SI}})(\text{LI}_{k} - \bar{\text{LI}})}{\sqrt{\sum_{k=1}^{K} (\text{SI}_{k} - \bar{\text{SI}})^2} \sqrt{\sum_{k=1}^{K} (\text{LI}_{k} - \bar{\text{LI}})^2}},
\end{equation}
where $\text{SI}_{k}$ and $\text{LI}_{k}$ denote Speech (SI) and Lip Intensity (LI), respectively, 
$\bar{\text{SI}} {=} \tfrac{1}{K}\sum_{k=1}^{K} \text{SI}_{k}$ and $\bar{\text{LI}} {=} \frac{1}{K}\sum_{k=1}^{K} \text{LI}_{k}$.
We define SI using speech loudness, specifically the z-normalized Root Mean Square (RMS) value, which is a widely accepted measure of speech intensity in signal processing.
To define LI, we measure the averaged lip displacement value of a video clip, followed by the z-normalization.

\section{Experiments}
\label{sec:experiments}

We first outline 
the evaluation setup, and
then present thorough analyses of the experimental results.
Due to the space limitation, we present more implementation details and additional experiments in the supplementary material.

\subsection{Experimental Settings}
\paragraph{Datasets}
Most of the existing speech-driven 3D talking head generation methods rely on VOCASET~\cite{voca} and BIWI~\cite{biwi} to train and test the models.
However, these datasets have limited ranges of facial motion patterns due to the small
dataset scale and restricted intensity range, which restricts their ability to fully capture the intricate relationship between speech and 3D face mesh.
To address the lack of training dataset, we construct two large-scale speech-3d face mesh benchmark datasets, LRS3-3D and MEAD-3D, by processing LRS3 and MEAD videos using two monocular face reconstruction methods: SPECTRE~\cite{spectre} for LRS3~\cite{lrs3}, which ensures accurate lip movements, and SMIRK~\cite{retsinas20243d} for MEAD~\cite{mead}, which captures diverse speech and lip movement intensities.
We use LRS3 in the first stage and LRS3-3D in the second stage to train our speech-mesh synchronized representation.
Then, for base model training and evaluations, we adopt two configurations: (1) training and testing with VOCASET, in line with existing work, and (2) combining MEAD-3D with VOCASET during training to endow expressiveness and testing on MEAD-3D.

\paragraph{Base methods}
We use three state-of-the-art 3D talking head generation models~\cite{faceformer,codetalker,selftalk} to evaluate the effectiveness of our perceptual loss.

\paragraph{Metrics}
To comprehensively evaluate the three aspects of lip synchronization, we assess MTM, PLRS, and SLCC, corresponding to temporal synchronization, lip readability, and expressiveness, respectively. 
Additionally, we compute the level-wise SLCC for the MEAD-3D test set, which includes three distinct emotional intensity levels, to evaluate the expressive capability of 3D talking head generation models. 
We also measure LVE as part of the lip readability evaluation.

\begin{table}[t]
\centering
    \resizebox{1.0\linewidth}{!}{
    \begin{tabular}{lcccl}
    \toprule
    \multirow{3}{*}{Method}  &  Temporal & \multicolumn{2}{c}{Lip} & \multirow{2}{*}{Expressiveness} \\
    &  Synchronization& \multicolumn{2}{c}{Readability} & \\
    \cmidrule(lr){2-2}\cmidrule(lr){3-4} \cmidrule(lr){5-5}
    &MTM ($\downarrow$) & LVE ($\downarrow$) & PLRS ($\uparrow$) & \text{SLCC} / $\Delta$ ($\downarrow$)\\
    \midrule
    VOCASET & - & - & 0.409 & 0.34 / -  \\ 
    \midrule
    FaceFormer~\cite{faceformer} & 53.6  & 3.357 & 0.368 & 0.26 / 0.08  \\ 
    \quad + Ours rep. & \textbf{52.2}  & \textbf{3.091} & \textbf{0.463} & 0.37 / \textbf{0.03} \\ 
    \midrule
    CodeTalker~\cite{codetalker} & 61.8 & 3.700 & 0.381 & 0.38 / 0.04 \\ 
    \quad + Ours rep. & \textbf{60.9}  & \textbf{3.579} & \textbf{0.388} & 0.35 / \textbf{0.01} \\  
    \midrule
    SelfTalk~\cite{selftalk} & 50.1 & 2.971 & 0.414 & 0.41 / 0.07 \\  
    \quad + Ours rep. & \textbf{49.2}  & \textbf{2.924} & \textbf{0.418} & 0.35 / \textbf{0.01} \\ 
    \bottomrule
        \end{tabular}
        }
        \caption{\textbf{Quantitative results of lip synchronization on VOCASET~\cite{voca} test set.} 
        We evaluate the base models on our proposed lip synchronization metrics.
        We denote $\Delta$ as the difference in SLCC between the model and those measured on the data distribution.
        A lower $\Delta$ indicates the model more closely represents the intensity correlation of the dataset.
        We demonstrate the effectiveness of our representation in consistently enhancing all three aspects of lip synchronization.
        }
        \label{table:eval_vocaset}
    \vspace{-5mm}
    \end{table}
    
\begin{figure}[t]
    \centering
    \includegraphics[width=\linewidth]{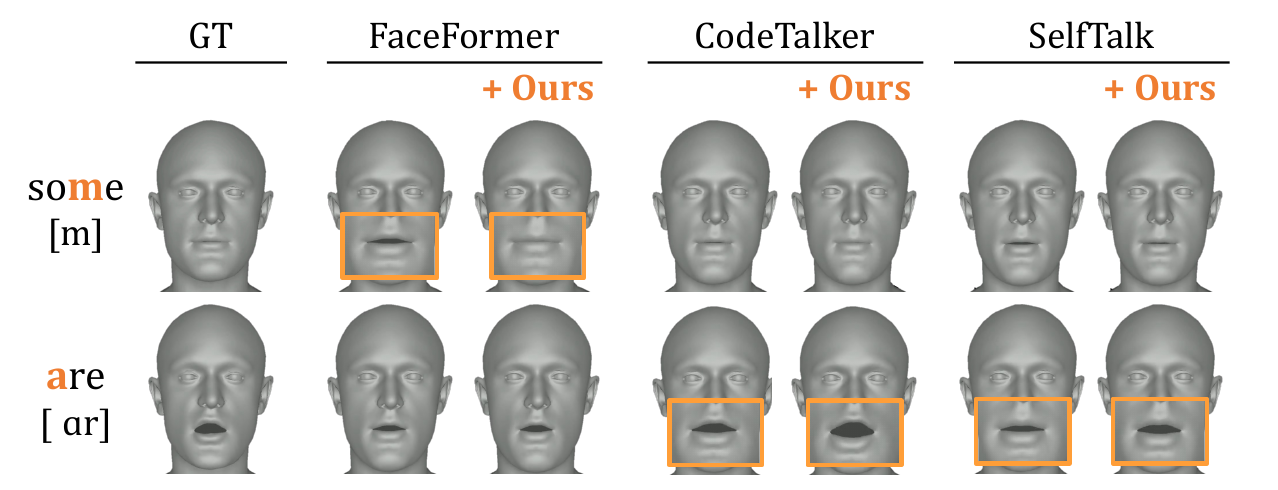}
    \vspace{-6mm}
    \caption{\textbf{Qualitative results of the effectiveness of our perceptual loss for lip readability.} Our perceptual loss guides baselines~\cite{faceformer,codetalker,selftalk} to generate perceptually accurate lip movements.}
    \label{fig:lip_readability_qualitative}
    \vspace{-4mm}
\end{figure}

\subsection{Experimental Results and Analysis}
We conduct evaluations to assess the effectiveness of our speech-mesh synchronized representation and the incorporation of an expressive speech-3D face mesh paired dataset (\ie, MEAD-3D) in enhancing the three criteria. 

\begin{table}[t]
\centering
\resizebox{0.95\linewidth}{!}{%
\Large
\begin{tabular}{llcccl}
\toprule
\multirow{3}{*}{Method}  & \multirow{3}{*}{Perceptual Loss }  & Temporal & \multicolumn{2}{c}{Lip} & \multirow{2}{*}{Expressiveness} \\
&&  Synchronization & \multicolumn{2}{c}{Readability} & \\
\cmidrule(lr){3-3}\cmidrule(lr){4-5} \cmidrule(lr){6-6}
&&MTM ($\downarrow$)& LVE ($\downarrow$)& PLRS ($\uparrow$) &SLCC / $\Delta$ ($\downarrow$)\\
\midrule
VOCASET & - & - & - & 0.409 & 0.34 / -  \\ 
\midrule
\multirow{4}{*}{FaceFormer~\cite{faceformer}}&\ding{55} & 53.6& 3.357 & 0.368 & 0.26 / 0.08   \\ 
&3D SyncNet & 55.6 & 3.316 & \underline{0.435}& 0.38 / \underline{0.04} \\ 
&Ours w/o 2D prior  & \underline{55.3} & \underline{3.278} & 0.400& 0.42 / 0.08  \\
&Ours w/ 2D prior & \textbf{52.2}  & \textbf{3.091} & \textbf{0.463} & 0.37 / \textbf{0.03} \\ 
\midrule
\multirow{4}{*}{CodeTalker~\cite{codetalker}}&\ding{55} & 61.8 & 3.700 & \underline{0.381} & 0.38 / \underline{0.04} \\ 
&3D SyncNet & \underline{59.6}  & 4.319 & 0.379& 0.14 / 0.20  \\ 
&Ours w/o 2D prior & \textbf{55.9} & \textbf{3.579} & 0.374& 0.23 / 0.11  \\
&Ours w/ 2D prior & 60.9  & \textbf{3.579} & \textbf{0.388} & 0.35 / \textbf{0.01} \\ 
\midrule
\multirow{4}{*}{SelfTalk~\cite{selftalk}}&\ding{55} & 50.1 & \underline{2.971} & 0.414 & 0.41 / 0.07  \\ 
&3D SyncNet & \underline{49.5} & 2.941 & 0.405& 0.35 / \textbf{0.01}  \\ 
&Ours w/o 2D prior  & 54.4 & 3.149 & \underline{0.417}& 0.39 / 0.05  \\
&Ours w/ 2D prior & \textbf{49.2}  & \textbf{2.924} & \textbf{0.418} & 0.35 / \textbf{0.01} \\ 
\bottomrule
\end{tabular}
}
    \caption{\textbf{Ablation study on architectural choice and 2D prior knowledge.} 
    We validate the effectiveness of the transformer-based architecture and curriculum learning with a pre-trained 2D speech representation by ablating them from our proposed representation.}
    \label{table:ablation}
    \vspace{-4mm}
\end{table}

\question{How well do existing 3D talking head models achieve lip synchronization in all three aspects?}
The results are summarized in~\Tref{table:eval_vocaset}.
For temporal synchronization, most base models achieve MTM values between 50 and 60ms, indicating performance close to the 
acceptable asynchrony threshold. 
Regarding lip readability, SelfTalk~\cite{selftalk} achieves the best performance, while FaceFormer~\cite{faceformer} has the lowest PLRS score. 
Furthermore, CodeTalker~\cite{codetalker} demonstrates the closest SLCC values to the ground truth VOCASET mesh, while FaceFormer exhibits the highest SLCC discrepancy.

\question{Does our speech-mesh representation improve lip synchronization?}
Yes. 
\Tref{table:eval_vocaset} and \Fref{fig:lip_readability_qualitative} show consistent improvements in the three aspects
with
our perceptual loss.

\begin{figure}[t]
    \centering
    \includegraphics[width=0.8\linewidth]{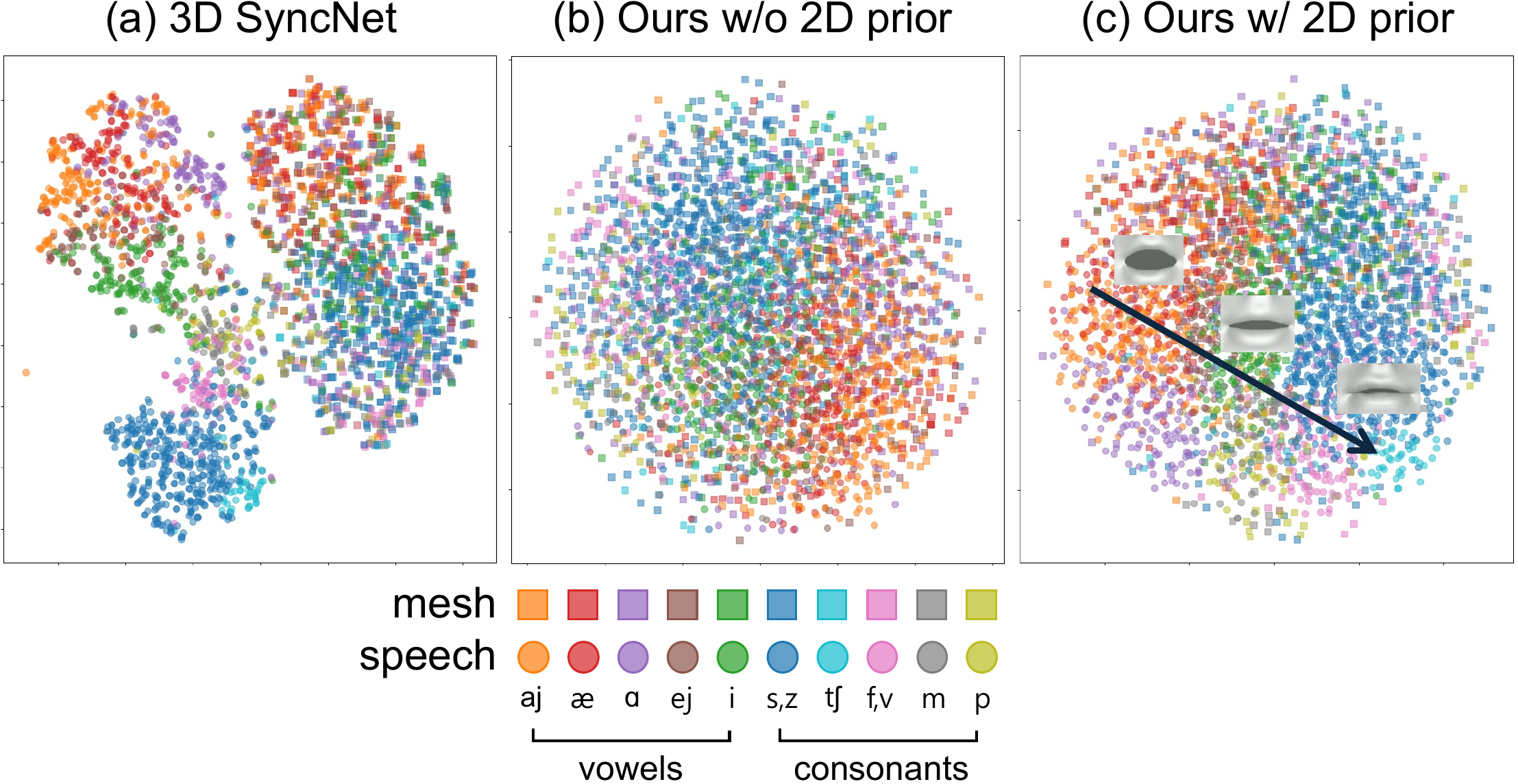}\vspace{-2mm}
    \caption{\textbf{t-SNE plot of ablation study.}
    We plot the t-SNE graph for each perceptual critic model.
    We represent the features with same phoneme as same color.
    Squared and circled points denote mesh and speech features from each representation, respectively.
    }
    \label{fig:tsne_ablation}
    \vspace{-3mm}
\end{figure}
\begin{figure}[t]
    \centering
    \includegraphics[width=0.8\linewidth]{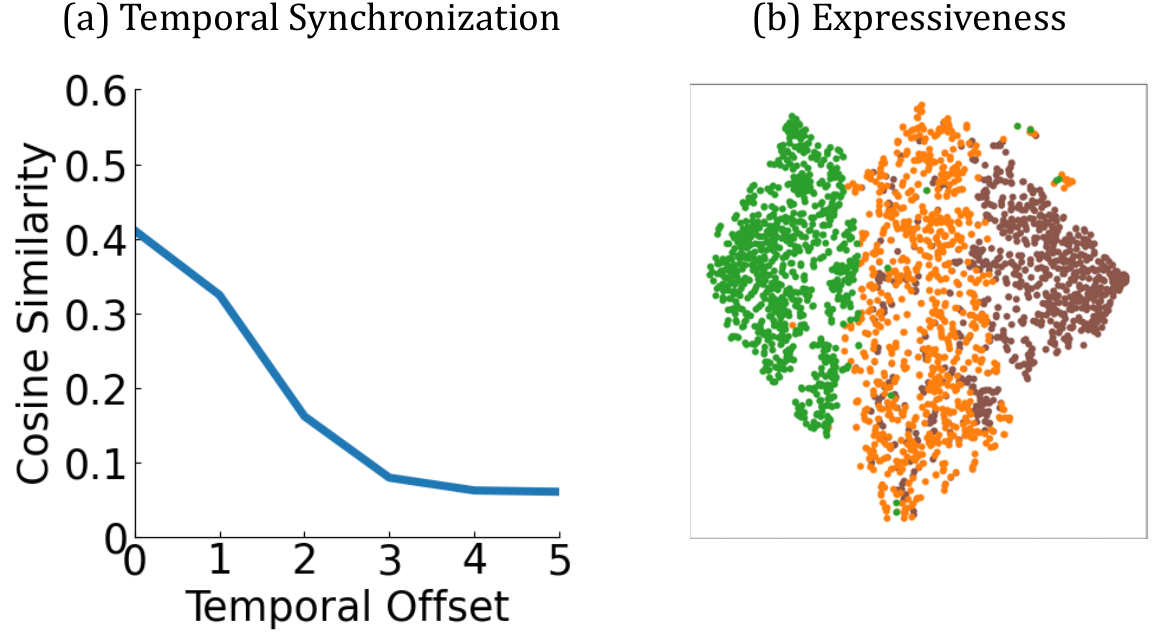}
    \vspace{-2mm}
    \caption{\textbf{Behaviors of our representation in temporal and expressiveness sensitivity.}
    We demonstrate the effectiveness of our representation in temporal synchronization and expressiveness using a cosine similarity graph and speech feature plots, respectively. We color the point as \textcolor{Brown}{low}, \textcolor{Orange}{medium}, and \textcolor{Green}{high} intensity.
    }
    \label{fig:ablation_model}
    \vspace{-4mm}
\end{figure}

\begin{figure*}[tp]
    \centering
    \includegraphics[width=0.85\linewidth]{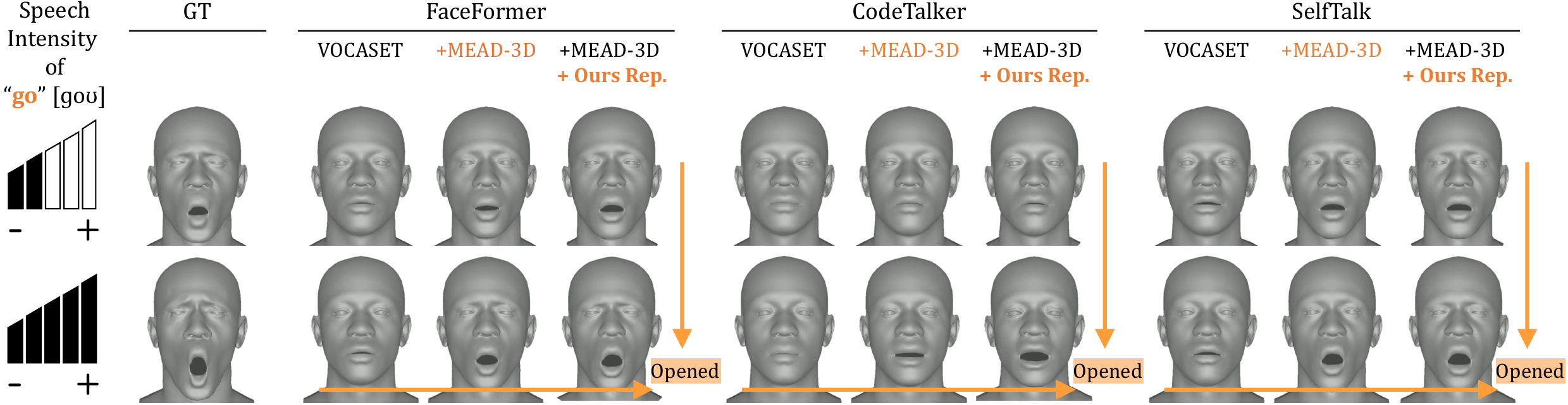}
    \vspace{-2mm}
    \caption{\textbf{Qualitative results for the expressiveness.} Given high and low intensity levels of speech, models trained on both MEAD-3D and VOCASET show more expressive lip movements compared to those trained on VOCASET alone, and even better with our perceptual loss.}
    \label{fig:expressive_qualitative}
    \vspace{-4mm}
\end{figure*}

\question{What makes our speech-mesh representation have lip synchronization ability?}
We hypothesize that the transformer-based architecture and curriculum learning with a pre-trained audio-visual speech representation contribute significantly to improved lip synchronization. 
To validate this, we conduct ablation studies, summarized in \Tref{table:ablation}, examining the roles of pre-trained speech representation and architectural design. 
Without the pre-trained speech representation, base models show 
lower performance across metrics, and CNN-based architectures, 3D SyncNet, do not clearly show effectiveness, while ours with 2D prior consistently show improvement.
The t-SNE plots of perceptual critic models in \Fref{fig:tsne_ablation} show that, in (a) 3D SyncNet, speech and mesh features corresponding to the same phoneme group are separated. 
In (b) ours without 2D priors, 
speech and mesh features lack separation and appear scattered without clustering. 
Our representation (c), notably, tends to form more distinct clusters 
according to 
phonemes, with vowels and consonants grouped closely, potentially contributing to enhancing lip readability.
We also observe a directional progression in the feature space, shifting from phonemes with mouth opening (\eg, /aj/) to those with mouth closing (\eg, /f/).

We also examine our representation's performance in other aspects,
temporal synchronization and expressiveness. 
Figure~\ref{fig:ablation_model}-(a) demonstrates temporal sensitivity, as cosine similarity drops when temporal misalignment is introduced between input speech and 3D face mesh. 
In \Fref{fig:ablation_model}-(b), we plot speech features at varying speech intensities, showing a directional trend as intensity increases from lowest to highest.
Figures~\ref{fig:tsne_ablation} and \ref{fig:ablation_model} imply that our representation holds favorable properties discussed in \Fref{fig:teaser} for the three criteria.

\question{Can we unlock the expressive power of 3D talking heads?} 
Likely. Since VOCASET~\cite{voca} lacks the range of diverse speech and lip movement intensities, evaluating expressive power requires testing on a dataset with a broader range of intensities. 
To study this, we examine the expressiveness of the base models on the MEAD-3D dataset, which includes three intensity levels. 
We assess SLCC at each intensity level to evaluate expressiveness and identify any expressiveness bounds as the level of intensity increases.
We first test the base models trained on VOCASET~\cite{voca} against the MEAD-3D test set. 
\Tref{table:mead3d} shows these models demonstrate limited expressiveness, with SLCC values showing minimal increase across intensity levels.
We hypothesize that this limitation arises, because VOCASET’s smaller scale and intensity range restrict the model’s ability to learn relationships between speech and lip intensity.
To address this, we integrate MEAD-3D with VOCASET for training, aiming to boost expressiveness in the base models. This approach consistently improves SLCC across all intensity levels. 
However, simply adding MEAD-3D degrades lip synchronization metrics, except for expressiveness, \ie, non-trivial.
To counterbalance this, we leverage our 
perceptual loss, which effectively mitigates the degradation introduced by MEAD-3D while improving
expressiveness
(see \Fref{fig:expressive_qualitative}).

\begin{table}[!t]
\centering
    \Large
    \resizebox{1.0\linewidth}{!}{
    \begin{tabular}{lcccllll}
    \toprule
    \multirow{3}{*}{Method}  &  Temporal & \multicolumn{2}{c}{Lip} & \multicolumn{4}{c}{\multirow{2}{*}{Expressiveness}} \\
    &  Synchronization& \multicolumn{2}{c}{Readability} & \\
    \cmidrule(lr){2-2}\cmidrule(lr){3-4}\cmidrule(lr){5-8}
    &\multirow{2}{*}{MTM ($\downarrow$)}  & \multirow{2}{*}{LVE ($\downarrow$)} & \multirow{2}{*}{PLRS ($\uparrow$)} &\multicolumn{4}{c}{SLCC / $\Delta$ ($\downarrow$)}\\
    & && & Lv1 & Lv2 & Lv3 & Avg \\
    \midrule
    MEAD-3D & - &  - & 0.230 & 0.24 / - & 0.30 / - & 0.39 / - & 0.42 / - \\
    \midrule
    FaceFormer~\cite{faceformer} & 59.6  & 3.207 & \underline{0.299} & 0.08 / 0.16 & 0.07 / 0.23 & 0.07 / 0.32 & 0.06 / 0.36   \\
    \quad + MEAD-3D & \underline{59.5} &\underline{1.139} &0.176 & 0.26 / \textbf{0.02} & 0.30 / \textbf{0.00} & 0.34 / \textbf{0.05} & 0.35 / \textbf{0.07}\\ 
    \quad\quad + Ours rep.  & \textbf{55.8} & \textbf{1.114} & \textbf{0.306} & 0.27 / \underline{0.03} & 0.27 / \underline{0.03} & 0.32 / \underline{0.07} & 0.33 / \underline{0.09}  \\ 
    \midrule
    CodeTalker~\cite{codetalker}  &  \underline{60.7} & 3.236 & \textbf{0.294}& 0.02 / 0.22 & 0.03 / 0.27 & 0.03 / 0.36 & 0.02 / 0.40  \\ 
    \quad + MEAD-3D & 60.9 & \underline{2.954} &0.154 & 0.09 / \underline{0.15} & 0.12 / \underline{0.18} & 0.06 / \underline{0.33} & 0.11 / \underline{0.31} \\
    \quad\quad + Ours rep.  & \textbf{58.6} & \textbf{2.705} & \underline{0.221}& 0.18 / \textbf{0.06} & 0.29 / \textbf{0.01} & 0.31 / \textbf{0.08} & 0.31 / \textbf{0.11}\\  
    \midrule
    SelfTalk~\cite{selftalk}  & \underline{53.4} & 3.396 & \textbf{0.294}& 0.14 / 0.10 & 0.14 / 0.16 & 0.17 / 0.22 & 0.15 / 0.27  \\ 
    \quad + MEAD-3D  & 54.2 &\underline{1.238} &0.216 & 0.16 / \underline{0.08} & 0.28 / \underline{0.02} & 0.32 / \underline{0.07} & 0.31 / \underline{0.11}\\
    \quad\quad + Ours rep.  & \textbf{52.7} &\textbf{1.192} & \underline{0.230}& 0.17 / \textbf{0.07} & 0.29 / \textbf{0.01} & 0.34 / \textbf{0.05} & 0.33 / \textbf{0.09} \\ 
    \bottomrule
        \end{tabular}
        }
        \caption{\textbf{Quantitative results of lip synchronization on MEAD-3D test set.}
        We evaluate the base models on the MEAD-3D test set.
        We also compute the level-wise SLCC to evaluate the expressive capability of the models.
        }
        \label{table:mead3d}
    \vspace{-8mm}
    \end{table}

\section{Conclusion}
\label{sec:conclusion}

This paper addresses challenges in existing 3D talking head generation models, which often overlook the true correspondence between speech and lip movements, making it difficult to accurately link lip movements with varying speech characteristics. To overcome this issue, we identify three essential aspects—Temporal Synchronization, Lip Readability, and Expressiveness—that influence the perceptual quality of lip movements, and develop specific metrics for each aspect. We introduce a speech-mesh synchronized representation that exhibits these emergent properties and adopt it as a perceptual loss.
Our extensive analyses
demonstrate that our perceptual loss consistently enhances models across three aspects. We believe that our defined aspects will guide future research in generating more 
realistic 
3D talking heads, and our representation will serve as a key component.

\clearpage

\paragraph{Acknowledgments}
This research was supported by a grant from KRAFTON AI, and was also partially supported by Institute of Information \& communications Technology Planning \& Evaluation (IITP) grant funded by the Korea government (MSIT) (No.RS-2021-II212068, Artificial Intelligence Innovation Hub; No.RS-2023-00225630, Development of Artificial Intelligence for Text-based 3D Movie Generation; No. RS-2024-00457882, National AI Research Lab Project) and Culture, Sports and Tourism R\&D Program through the Korea Creative Content Agency grant funded by the Ministry of Culture, Sports and Tourism in 2024 (Project Name: Development of barrier-free experiential XR contents technology to improve accessibility to online activities for the physically disabled, Project Number: RS-2024-00396700, Contribution Rate: 25\%).

{
    \small
    \bibliographystyle{ieeenat_fullname}
    \bibliography{main}
}

\clearpage
\maketitlesupplementary

\newtheorem{assume}{Assumption}

\setcounter{section}{0}
\setcounter{figure}{0}
\setcounter{table}{0}
\setcounter{equation}{0}

\renewcommand{\thesection}{\Alph{section}}
\renewcommand{\thefigure}{S\arabic{figure}}
\renewcommand{\thetable}{S\arabic{table}}
\renewcommand{\theequation}{\alph{equation}}

\hypersetup{linkcolor=black}

\section*{Contents}

\noindent\hyperref[sec:A]{\textbf{A \ \ \ Supplementary Video}}\\
\hyperref[sec:B]{\textbf{B \ \ \ Emergent Properties of 2D Speech Representation}} \\ 
\hyperref[sec:C]{\textbf{C \ \ \ Speech-Mesh Synchronized Representation}}\\
\hyperref[sec:C.1]{\text{\qquad C.1 \ \ \ Network architecture}} \\ 
\hyperref[sec:C.2]{\text{\qquad C.2 \ \ \ Training pipeline}} \\ 
\hyperref[sec:C.3]{\text{\qquad C.3 \ \ \ Dataset statistics}} \\
\hyperref[sec:D]{\textbf{D \ \ \ Details of Human Study on Lip Synchronization Criteria}}\\
\hyperref[sec:E]{\textbf{E \ \ \ Evaluation Metrics}} \\
\hyperref[sec:E.1]{\text{\qquad E.1 \ \ \ 
Definition and implementation details
}} \\ 
\hyperref[sec:E.2]{\text{\qquad E.2 \ \ \ 
Human study on perceptual metric
}} \\
\hyperref[sec:F]{\textbf{F \ \ \ Implementation Details of Ablation Study}}\\
\hyperref[sec:G]{\textbf{G \ \ \ Additional Results}}\\
\hyperref[sec:G.1]{\text{\qquad G.1 \ \ \ 
Human study on applying perceptual loss
}} \\ 
\hyperref[sec:G.2]{\text{\qquad G.2 \ \ \ 
FDD evaluation on applying perceptual loss}} \\ 
\hyperref[sec:G.3]{\text{\qquad G.3 \ \ \ 
Qualitative result of temporal synchronization
}} \\ 
\hyperref[sec:G.4]{\text{\qquad G.4 \ \ \ 
Stability comparison on loss and cosine similarity
}} \\ 
\noindent\hyperref[sec:H]{\textbf{H \ \ \ Discussion}}\\

\noindent\rule{\linewidth}{0.2pt}

\appendix

\hypersetup{linkcolor=cvprblue}

\section{Supplementary Video}\label{sec:A}
This work focuses on 3D facial motions, which are best viewed in video format. 
Please refer to the attached \textbf{supplementary video}.
The video contains qualitative results of lip synchronization on the VOCASET and MEAD-3D test sets, demonstrating the effectiveness of our method in enhancing lip synchronization in aspects of lip readability and expressiveness.

\section{Emergent Properties of 2D Speech Representation}\label{sec:B}

\begin{figure}[t]
    \centering
    \includegraphics[width=1\linewidth]{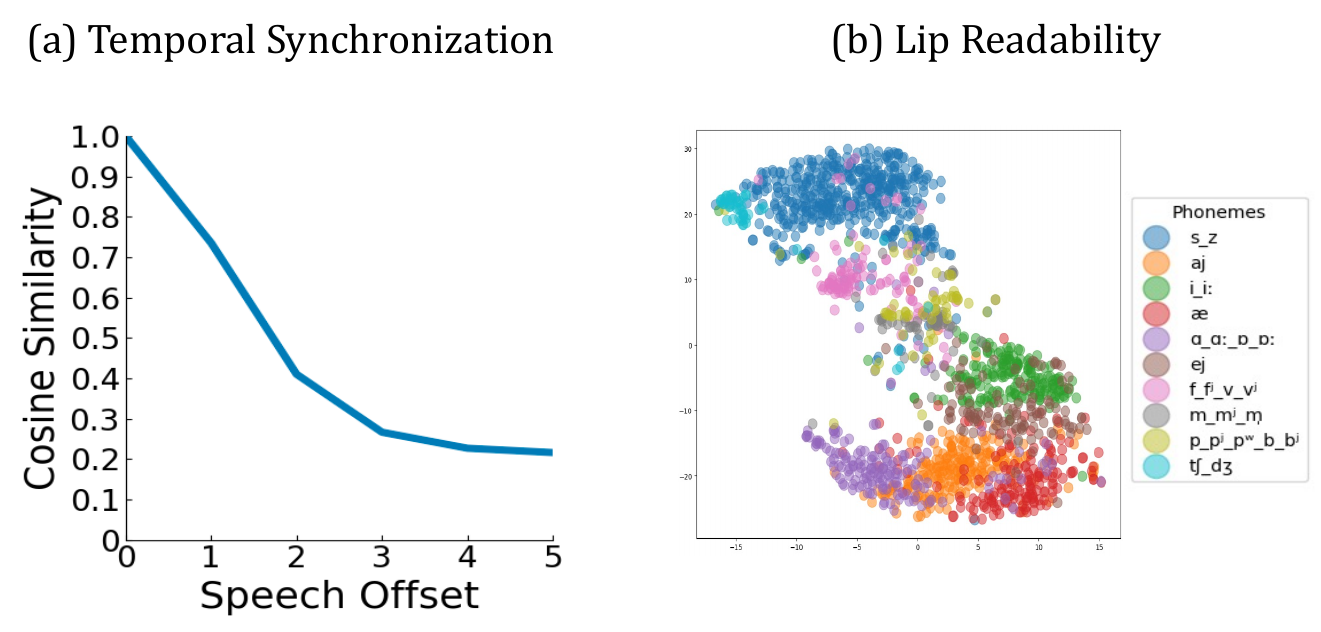}
    \caption{\textbf{Emergent properties of 2D speech representation.}
    We visualize a cosine similarity versus temporal offset graph and a t-SNE visualization of the 2D audio-visual speech representation.
    The 2D speech representation already possesses desired properties we pursue, which motivates us to transfer the emergent properties to the speech-mesh representation space. 
    }
    \label{fig:model_capacity}
\end{figure}
In this section, we conduct further analyses of 2D speech representation (\ie, 2D prior knowledge), which motivate the transfer of the emergent properties of 2D speech representation to the speech-mesh representation space using a curriculum learning approach.

We observe that the 2D audio-visual speech representation, trained with a transformer architecture and an extensive video dataset~\cite{lrs3}, inherently exhibits the desirable properties for lip synchronization that we aim to achieve.
We visualize a cosine similarity versus temporal offset graph and a t-SNE visualization of the 2D audio-visual speech representation in~\Fref{fig:model_capacity}.
The speech representation exhibits the properties regarding the critical aspects of lip synchronization: (1) Temporal sensitivity in~\Fref{fig:model_capacity}-(a), (2) clear separation and clustering of speech features corresponding to the same phoneme group in~\Fref{fig:model_capacity}-(b), and (3) a directional progression of speech features as intensity increases from the lowest to the highest levels in Fig.~5-(b) of the main paper~\footnote{We freeze the pre-trained speech encoder from stage 1 and utilize it as the speech encoder in stage 2, which ensures that the speech representation in both stages shares the same favorable property of expressiveness.}.
This motivates us to transfer these emergent properties to the 3D speech-mesh representation through the curriculum learning approach, as mentioned in Sec.~4 of the main paper.
Furthermore, as shown in Figs.~4 and 5 of the main paper, we demonstrate that these properties are successfully transferred to the speech-mesh representation.

\section{Speech-Mesh Synchronized Representation}\label{sec:C}
We provide more details on the network architecture of audio-visual speech representation and speech-mesh representation (Sec.~\ref{sec:C.1}).
In addition, we provide the training details of the two-stage training process (Sec.~\ref{sec:C.2}) and dataset statistics of speech-mesh benchmark datasets (Sec.~\ref{sec:C.3}).
\subsection{Network architecture}\label{sec:C.1}
\begin{table*}[t]
\centering
    \resizebox{1.0\linewidth}{!}{\arraybackslash
    \begin{tabular}{c|c|l|l}
    \toprule
    Stage &Module & Input $\rightarrow$ Output& Layer Operation  \\ 
    \cmidrule(lr){1-4}
    \multirow{16}{*}{1}&Speech Tokenizer & $\mathbf{X_s}(C_s,H_s,W_s) \rightarrow \mathbf{S}(N,H)$ & Conv2D$((1,16),(1,16),H)$   \\ 
    \cmidrule(lr){2-4}
    &\multirow{2}{*}{Speech Encoder} & $\mathbf{S}^{unmask}(N^{unmask},H) \rightarrow$  & \multirow{2}{*}{$[$MHSA$(H,8) \rightarrow$ FFN$(H)] \times 10 \rightarrow$ LN}  \\ 
    && $ \mathbf{Z_s}(N^{unmask},H)$  &   \\ 
    \cmidrule(lr){2-4}
    &\multirow{4}{*}{Speech Decoder} & \multirow{3}{*}{$\mathbf{F_s'} \rightarrow \hat{\mathbf{S}}(N^{mask},C_s \cdot H_s\cdot W_s)$}  & Concat$($Linear$(H,384) +$ PE$(N^{unmask})$, PE$(N^{mask})) \rightarrow$\\
    &&  &MHSA$(384,8) \rightarrow$ FFN$(384 \cdot 4) \rightarrow$  \\
    &&   &$[$MHSA$(384,8) \rightarrow$ MHCA$(Z_s,384,6) \rightarrow$ FFN$(384 \cdot 4)] \times 3 \rightarrow$ \\
    &&   &LN $\rightarrow$ Linear$(C_s \cdot H_s\cdot W_s) \rightarrow$ Slice$[N^{unmask}:]$ \\
    \cmidrule(lr){2-4}
    &Video Tokenizer & $\mathbf{X_v}(C_v,T,H_v,W_v) \rightarrow \mathbf{V}(M,H)$  & Conv3D$\big((1, 16, 16),(1, 16, 16),H\big)$  \\
    \cmidrule(lr){2-4}
    &\multirow{2}{*}{Video Encoder} & $\mathbf{V}^{unmask}(M^{unmask},H)\rightarrow$  & \multirow{2}{*}{$[$MHSA$(H,8) \rightarrow$ FFN$(H)] \times 10 \rightarrow$ LN}  \\
    & & $\mathbf{Z_v}(M^{unmask},H)$  &  \\
    \cmidrule(lr){2-4}
    &\multirow{4}{*}{Video Decoder} & \multirow{3}{*}{$\mathbf{F_v'} \rightarrow \hat{\mathbf{V}}(M^{mask},C_v \cdot H_v\cdot W_v)$}  & Concat(Linear$(H,384) +$ PE$(M^{unmask})$, PE$(M^{mask})) \rightarrow$\\
    &&  &MHSA$(H,8) \rightarrow$ FFN$(H) \rightarrow$  \\
    &&   &$[$MHSA$(H,8) \rightarrow$ MHCA$(Z_v,H,6) \rightarrow$ FFN$(H)] \times 3 \rightarrow$  \\
    &&   &LN $\rightarrow$ Linear$(C_v \cdot H_v\cdot W_v) \rightarrow$ Slice$[M^{unmask}:]$\\
    \cmidrule(lr){2-4}
    &\multirow{2}{*}{Fusion Encoder} & $\mathbf{Z_s}, \mathbf{Z_v} \rightarrow \mathbf{F_s}(N^{unmask},H)$ & $[$MHSA$(H,8) \rightarrow$ MHCA$(Z_v,H,8) \rightarrow$ FFN$(H \cdot 4)] \times 2$  \\
    && $\mathbf{Z_s}, \mathbf{Z_v} \rightarrow \mathbf{F_v}(M^{unmask},H)$ & $[$MHSA$(H,8) \rightarrow$ MHCA$(Z_s,H,8) \rightarrow$ FFN$(H \cdot 4)] \times 2$ \\
    \midrule
    \multirow{2}{*}{2}&Mesh Tokenizer & $\mathbf{X_m}(T,V \cdot 3) \rightarrow \mathbf{M}(T,H)$ & Linear$(H)$   \\
    \cmidrule(lr){2-4}
    &Mesh Encoder & $\mathbf{M} \rightarrow \mathbf{Z_m}(T,H)$  & $[$MHSA$(H,8) \rightarrow$ FFN$(H)] \times 10 \rightarrow$ LN  \\
    \bottomrule
    \end{tabular}
    }
    \caption{\textbf{Architecture details.} 
    The parameters of network architectures. 
    Conv2D$(k,s,n)$ denotes a 2D Convolutional layer with kernel size $k$, stride size $s$, and output channel of $n$. 
    MHSA$(d,nhead)$ denotes a multi-head self-attention layer with the input channels $d$ and the number of heads in multi-head attention $nhead$. 
    MHCA$(ca,d,nhead)$ denotes a multi-head cross-attention layer with additional cross-attention input $ca$. 
    PE$(a)$ is a position embedding layer where $a$ denotes the length of the position vector. 
    FFN$(d)$ is a feed-forward layer. 
    Linear$(n)$ denotes a linear layer with output channels of $n$. 
    LN denotes layer normalization and Slice$[s:]$ denotes slice operation.
    }
    \label{table:network_architecture}
\end{table*}

To improve the reproducibility of our speech-mesh representation, we further illustrate the detailed network architectures for the audio-visual speech representation and the speech-mesh representation, which are shown in \Tref{table:network_architecture}.
\subsection{Training pipeline}\label{sec:C.2}

\paragraph{Two-stage training process}
In our experiment, we set $T$ = 5, $H$ = 512, and $P$ = 30.
For training the audio-visual speech representation, we use $C_s$ = 1, $H_s$ = 64, $W_s$ = 128, $N$ = 512 for speech modality and $C_s$ = 3, $H_v$ = 160, $W_v$ = 160, $M$ = 500 for video modality. We train the audio-visual speech representation using LRS on two NVIDIA A6000 for 100 epochs with the AdamW optimizer ($\beta_1$ = 0.9, $\beta_2$ = 0.95 and $\epsilon$ = 1$e$-8), where the learning rate is initialized as 3$e$-4, and the mini-batch size is set as 40. 
For training the speech-mesh representation, we use the number of vertices $V$ = 5023. We train the speech-mesh representation using LRS-3D with the mini-batch size of 80, and other hyper-parameters remain unchanged as Stage 1.

\paragraph{Perceptual loss}
We employ our speech-mesh representation as a perceptual loss to enhance the perceptual accuracy of the 3D talking head model. We finetune our speech-mesh representation using the VOCASET~\cite{voca} train split on an NVIDIA A6000 for 5 epochs with the initial learning rate 1$e$-4 and other hyper-parameters remain unchanged as Stage 2. 
To train the 3D talking head models with our perceptual loss, we split the generated mesh from the model into 5 frames using a sliding window size of 1. We make a batch of size 80 and get uni-modal embeddings from our representation. We additionally apply the InfoNCE loss with a weight of 1$e$-7 to the original training loss of the model.

\subsection{Dataset statistics}\label{sec:C.3}
\begin{table}[t]
    \centering
        \renewcommand{\arraystretch}{1.05} 
        \resizebox{.97\linewidth}{!}{
            \begin{tabular}{l c c c c c}
            \toprule
            Dataset & $\#$ Vertex clips & $\#$ Speaker IDs & Total hours & FPS & \\
            \midrule
             VOCASET & 475 & 12 & 0.5 & 30 &\\
             BIWI & 1109 & 14 & 1.4 & 25\\
             LRS3-3D & 17752 & 788 & 61.1 & 25 &\\  
             MEAD-3D & 8765 & 15 & 10.2 & 30 &\\     
            \bottomrule
            \end{tabular}}
    \caption{\textbf{Statistics of speech-mesh paired benchmark.} 
    We use VOCASET, LRS3-3D and MEAD-3D speech-mesh paired datasets in our experiments.
    We construct two large-scale speech-mesh benchmark datasets, LRS3-3D and MEAD-3D, using monocular face reconstruction methods.}
    \label{tab:statistics}
\end{table}

\begin{figure}[t]
    \centering
    \vspace{-2mm}
    \includegraphics[width=1.0\linewidth]{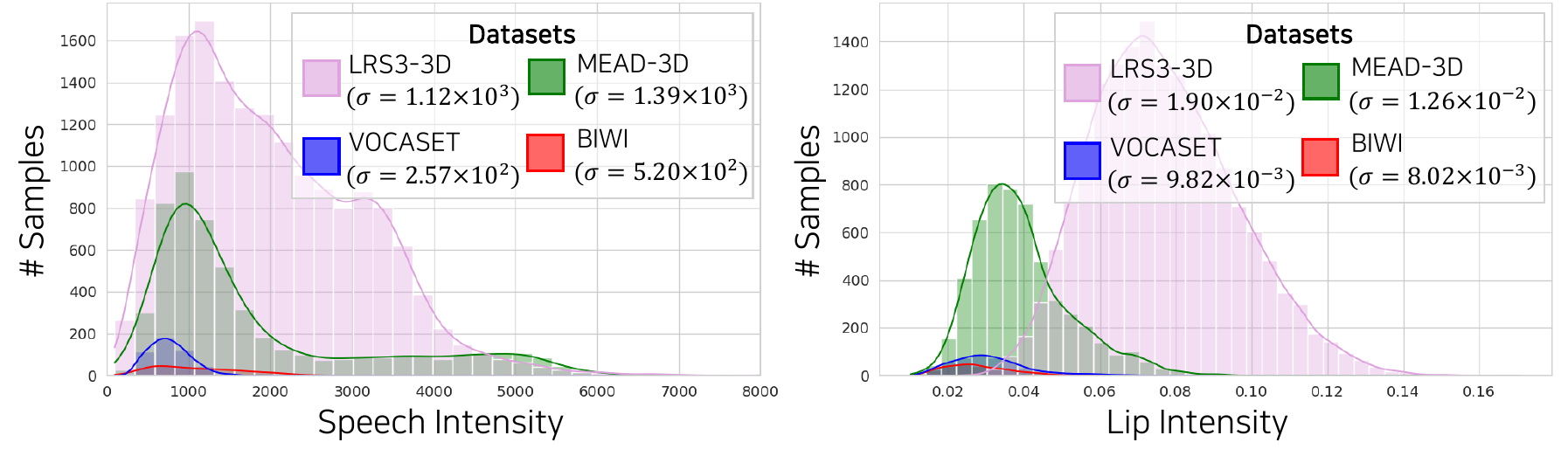}\vspace{-4mm}
    \caption{\textbf{Speech and lip intensity distributions across datasets.
    }We present speech and lip intensity distributions and corresponding standard deviation values across datasets.  
    }
    \label{fig:pattern}
    \vspace{-4mm}
\end{figure}

We construct LRS3-3D and MEAD-3D by processing LRS3~\cite{lrs3} and MEAD~\cite{mead} videos using two monocular face reconstruction methods, respectively: SPECTRE~\cite{spectre} for LRS3, which ensures accurate lip movements, and SMIRK~\cite{retsinas20243d} for MEAD, which captures diverse speech and lip movement intensities.
We construct a test split for LRS-3D, involving 934 clips.
We split MEAD-3D to construct a test split, which includes 3470 clips.

\Tref{tab:statistics} and Fig.~\ref{fig:pattern} show the statistics of the existing (VOCASET~\cite{voca}, BIWI~\cite{biwi}) and the newly proposed large-scale speech-mesh benchmark datasets (LRS3-3D and MEAD-3D).
As shown in \Tref{tab:statistics}, LRS3-3D and MEAD-3D have notably larger data sizes than VOCASET and BIWI.
Fig.~\ref{fig:pattern} presents the broader speech and lip intensity\footnote{
Lip intensity was normalized by eye distance to account for differences between FLAME and BIWI topologies.} distributions of LRS3-3D and MEAD-3D with higher standard deviations ($\sigma$), indicating greater variability in facial motions.
In contrast, VOCASET and BIWI show limitations in both scale and diversity.

\section{Details for Human Study on Lip Synchronization Criteria}\label{sec:D}
\paragraph{Human preference between the speech and lip intensities}
We conduct a preliminary experiment to demonstrate the positive correlation of human preference between the intensity of speech and lip movements in the 3D talking face field.
Using the intensity annotations from the MEAD dataset~\cite{mead}, we first split the MEAD-3D dataset into three categories: Level 1, Level 2, and Level 3, representing different intensity levels.
Then, we train a 3D talking face model~\cite{faceformer} using VOCASET~\cite{voca} (to ensure the quality of generation) and each intensity split separately.
This results in three distinct models, each of which tends to generate lip movements biased toward the intensity level present in its training data, regardless of the speech intensity provided as input.
We input three speeches with intensity levels ranging from Level 1 to Level 3 into each of the three biased models, producing nine intensity configurations in the generated mesh sequences as shown in Tab.1-[Left] of the main paper.
We then asked 17 participants, a balanced group of males and females from a non-expert background in the field, to rank their preferences in three videos, assigning a score from 1 (least preferred) to 3 (most preferred).
Each video has the same speech (identical in utterance and intensity) but differs in the intensity of the lip movements.

\paragraph{Human preference on Temporal sync. vs. Expressiveness}
We design a simple A/B test to investigate an interesting aspect of human perception for lip synchronization.
We use the two biased models from the previous human study: one trained to generate Level 1 lip movements and the other trained to generate Level 3 lip movements, regardless of the speech intensity.
For each model, we create two types of samples.
Sample A is temporally synchronized but lacks expressive synchronization (\eg, speech of Level 3 intensity and lip movements of Level 1 intensity).
In contrast, sample B has expressive synchronization (\eg, speech of Level 3 intensity and lip movements of Level 3 intensity) but is temporally misaligned.
To introduce the temporal mismatch in Sample B, we make the speech lead the lip movements by 100ms, which exceeds twice the established maximum acceptable synchrony~\cite{vatakis2006audiovisual}.
We then asked 28 participants, comprising a balanced group of males and females from a non-expert background in the field, to choose which sample they prefer based on how well the lip movements correspond to the speech in sample A vs. B.

\section{Evaluation Metrics}\label{sec:E}
We present the comprehensive definitions of the evaluation metrics and their implementation details (Sec.~\ref{sec:E.1}).
In addition, we provide the human study on the perceptual metric (Sec.~\ref{sec:E.2}), which demonstrates the correlation between our perceptual metric and human preference.

\subsection{Definition and implementation details}\label{sec:E.1}
\begin{figure}[t]
    \centering
    \includegraphics[width=1\linewidth]{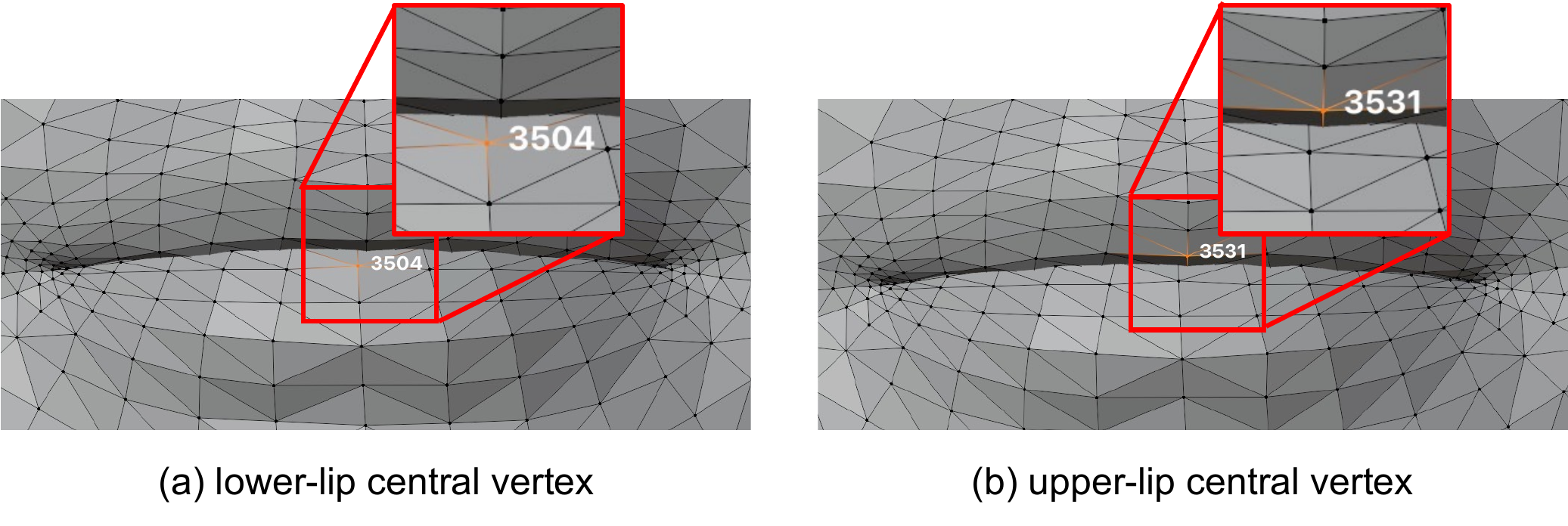}
    \caption{\textbf{Central vertices of the lower and upper lips.}
    We select two specific vertices that correspond to the center of the upper and lower lips to extract the lip vertex displacement sequences.
    }
    \label{fig:lips}
\end{figure}

\paragraph{Mean Temporal Misalignment (MTM)}
Let $\mathbf{V}(t)$ represent the ground truth vertex sequences, where each frame $t$ consists of vertex positions $\mathbf{v}_t \in \mathbb{R}^{N \times 3}$, with $N$ being the number of vertices. Similarly, $\mathbf{\hat{V}}(t)$ represents the predicted vertex sequences, with predicted vertex positions $\mathbf{\hat{v}}_t \in \mathbb{R}^{N \times 3}$. For each sample $k$, we select two specific vertices that correspond to the center of the upper and lower lips, extracting the upper-lip vertex sequence $\mathbf{V}_u(t) \in \mathbb{R}^{T \times 3}$ and the lower-lip vertex sequence $\mathbf{V}_l(t) \in \mathbb{R}^{T \times 3}$ (refer to Fig.~\ref{fig:lips}).

We then calculate the Euclidean distance between the upper and lower lip vertices over time to derive the ground truth lip distance sequence $d_v(t) = \left\| \mathbf{V}_u(t) - \mathbf{V}_l(t) \right\|$. The same process is applied to obtain the predicted lip distance sequence $\hat{d}_v(t)$. To reduce noise, we apply a Gaussian filter to both lip distance sequences.

\begin{figure}[t]
    \centering
    \includegraphics[width=1\linewidth]{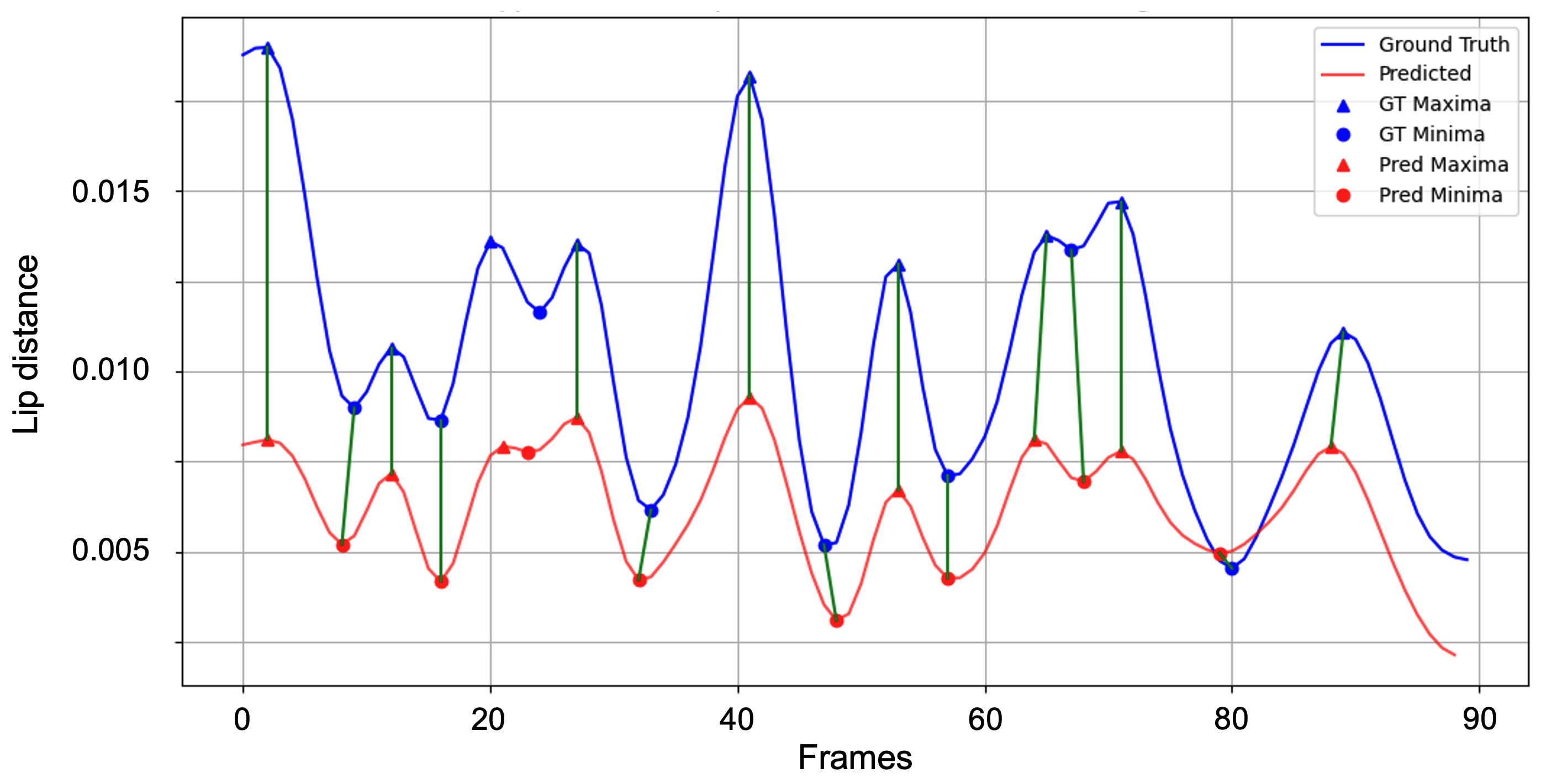}
    \caption{\textbf{An example of DDTW matching results between ground truth and predicted lip distance sequences.}
    We present an example of the DDTW local extrema correspondences
    of the ground truth and predicted lip vertex displacement sequences.
    We represent matched local extrema using green lines.
    }
    \label{fig:ddtw}
\end{figure}

Next, we compute the first-order derivatives of the smoothed lip distance sequences to capture the dynamic changes in lip movement. We then use Derivative Dynamic Time Warping (DDTW)~\cite{keogh2001derivative} to determine the optimal alignment path $\mathcal{A} = \{(i, j)\}$ between the derivative sequences $\delta \tilde{d}_v(t)$ and $\delta \tilde{\hat{d}}_v(t)$. We identify local extrema (peaks and valleys) in each derivative sequence and match only extrema of the same type (i.e., both maxima or both minima) to compute the absolute time difference $\delta t_n = |i - j|$ (refer to Fig.~\ref{fig:ddtw}).

\begin{figure}[t]
    \centering
    \includegraphics[width=1\linewidth]{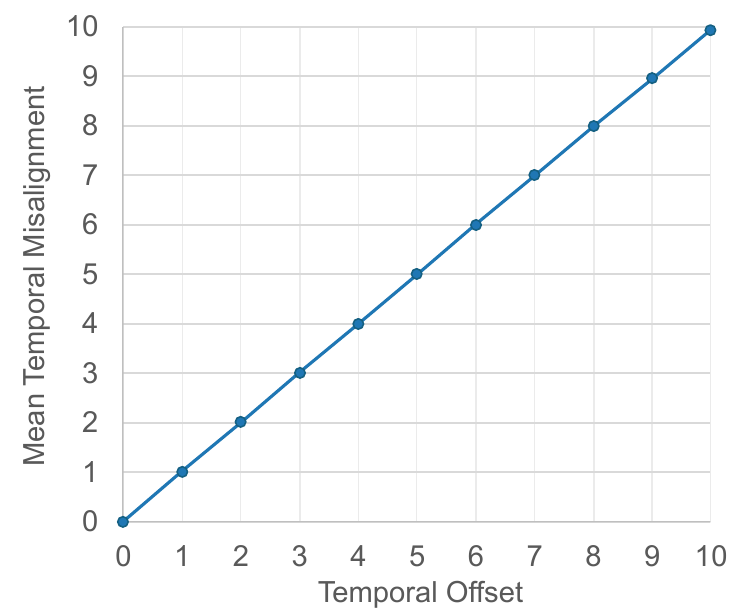}
    \caption{\textbf{Physical accuracy of Mean Temporal Misalignment.}
    We introduce temporal mismatch to the ground truth mesh sequences of VOCASET~\cite{voca} by shifting the speech to lead the mesh sequences by 0 to 10 frames (where 0 represents no mismatch).
    For each temporal offset, we calculate the average MTM and plot a graph showing the relationship between the temporal offset and the corresponding MTM values.
    }
    \label{fig:MTM_physical}
\end{figure}

For each sample $k$, the sample mean temporal misalignment $\Delta t_k$ is computed as $\Delta t_k = \dfrac{1}{M} \sum_{m=1}^{M} \delta t_n$, where $M$ is the number of matched extrema pairs in the sample. Finally, the overall mean temporal misalignment is given by $\overline{\Delta t} = \dfrac{1}{K} \sum_{k=1}^{K} \Delta t_k$, where $K$ is the total number of samples. A smaller $\overline{\Delta t}$ indicates better temporal alignment of the predicted sequences with the ground truth lip movements.
To express the Mean Temporal Misalignment (MTM) in milliseconds, we multiply $\overline{\Delta t}$ by the frame duration for the given dataset.
For instance, for a dataset with 25 FPS, the MTM is obtained by multiplying $\overline{\Delta t}$ by 40ms.
Refer to Algorithm~\ref{alg:temporal_misalignment} for more details on the MTM calculation.
Furthermore, to validate the physical accuracy of our proposed temporal synchronization metric, we present a graph showing the relationship between the temporal offset and the corresponding MTM values.
Specifically, we introduce temporal mismatch to the ground truth mesh sequences of VOCASET~\cite{voca} by making the speech leading the mesh sequences by 0 to 10 frames (\ie, 0 to 333ms for VOCASET).
Figure~\ref{fig:MTM_physical} shows that MTM accurately captures the degree of temporal mismatch across the samples, demonstrating the effectiveness and physical accuracy of our proposed temporal synchronization metric.

\paragraph{Perceptual Lip Readability Score (PLRS)}
We train speech-mesh representation using our proposed two-stage training process with different datasets, initializations, and batch sizes. For both Stage 1 and Stage 2, we use a batch size of 256. Given a speech and generated mesh pair $(\mathbf{X_s}, \hat{\mathbf{X}}_\mathbf{m})$, we split the generated mesh into 5 frames with a sliding window size of 5 to make mesh tokens $\{\hat{\mathbf{M}}_i\}_{i=1}^{G}$
, and the speech is also converted into corresponding speech tokens $\{\mathbf{S}_i\}_{i=1}^{G}$. We then compute the average cosine similarity between mean pooled speech embeddings $\{\mathbf{c}_{\mathbf{s},i}\}_{i=1}^{G}$ and mesh embeddings $\{\mathbf{c}_{\mathbf{m},i}\}_{i=1}^{G}$:
\begin{equation}
\label{eq:cos_sim}
{
PLRS(\mathbf{S},\hat{\mathbf{M}})=\frac{1}{G}\sum_{i=1}^{G}\frac{\mathbf{c}_{\mathbf{s},i}\cdot \mathbf{c}_{\mathbf{m},i}}{\left\| \mathbf{c}_{\mathbf{s},i}\right\|\left\| \mathbf{c}_{\mathbf{m},i}\right\|}.}
\end{equation}

\paragraph{Speech-Lip Intensity Correlation Coefficient (SLCC)}
First, we define speech intensity using speech loudness, specifically the Root Mean Square (RMS) value, which is a widely accepted measure of speech intensity in signal processing. RMS loudness effectively captures the energy of the speech signal and provides an accurate representation of perceived speech intensity.
However, since RMS values can vary based on recording conditions (\eg, microphone gain and distance from the microphone), we perform identity-wise z-normalization on the RMS values to standardize them, assuming that clips belonging to the same identity are recorded under similar conditions.
The Speech Intensity (SI) is thus defined as:
\begin{equation}
\label{eq:SI}
\text{SI}_{k} = \frac{\text{RMS}_k - \mu_{s,i}}{\sigma_{s,i}},
\end{equation}
where $\text{RMS}_k$ is the averaged RMS value of k-th video clip and $\mu_{s,i}$ and $\sigma_{s,i}$ are the mean and standard deviation of the speech RMS values for the clips with identity $i \in I$.

To define Lip Intensity (LI), we first measure the averaged lip displacement value of k-th video clip $\text{Dist}_k$.
as:
\begin{equation}
\label{eq:Dist}
\text{Dist}_k = \sqrt{\frac{1}{T_{k}-1} \sum_{t=1}^{T_{k}-1} \left( \frac{1}{V_l} \sum_{v=1}^{V_l} \| \mathbf{l}_{t+1, v} - \mathbf{l}_{t, v} \| \right)^2},
\end{equation}
where $T_k$ is the number of frames in clip $k$, $V_l$ is the number of vertices in the lip region, and $\mathbf{l}_{t, v} \in \mathbb{R}^{3}$ represents a vertex position in the lip region at time $t$.
Similar to Speech Intensity, we perform identity-wise z-normalization to the lip displacement values to mitigate individual bias in lip movement as:
\begin{equation}
\label{eq:LI}
\text{LI}_{k} = \frac{\text{Dist}_k - \mu_{l,i}}{\sigma_{l,i}},
\end{equation}
where $\mu_{l,i}$ and $\sigma_{l,i}$ are the mean and standard deviation of the lip displacement values for the clips belonging to identity $i \in I$.

Finally, we can obtain the Speech and Lip Correlation Coefficient as:
\begin{equation}
\label{eq:LI}
r_{SL} = \frac{\sum_{k=1}^{K} (SI_k - \bar{SI})(LI_k - \bar{LI})}{\sqrt{\sum_{k=1}^{K} (SI_k - \bar{SI})^2} \sqrt{\sum_{k=1}^{K} (LI_k - \bar{LI})^2}},
\end{equation}
where $\bar{SI} = \frac{1}{K}\sum_{k=1}^{K} SI_{k}$ and $\bar{LI} = \frac{1}{K}\sum_{k=1}^{K} LI_{k}$.

\subsection{Human study on perceptual metric}\label{sec:E.2}
To validate that our proposed perceptual metric, Perceptual Lip Readability Score (PLRS), effectively evaluates perceptual alignment, we conduct a human study that assesses the correlation between the metric scores and human preferences.
We collect meshes from the ground-truth VOCASET~\cite{voca} dataset and those generated by FaceFormer~\cite{faceformer}, CodeTalker~\cite{codetalker} and SelfTalk~\cite{selftalk}.
We measure the PLRS 
and the existing evaluation metric Lip Vertex Error (LVE)
for the generated meshes of each model, and subsequently rank the models by their PLRSs and LVEs.
We ask 16 participants, evenly balanced in gender and from non-expert backgrounds, to rank the models based on their preferences.
We then compute the Spearman's correlation coefficient $\rho$ to compare the PLRS rankings and the LVE rankings with the human preference rankings.
As shown in~\Tref{tab:PLRS}, PLRS exhibits a far more positive correlation with human preferences compared to the LVE.
This highlights the efficacy of our proposed metric in evaluating perceptual lip readability from a human perspective.

\begin{table}[t]
\centering
    \resizebox{0.5\linewidth}{!}{
    \begin{tabular}{cc}
        \hline
        Metric & Spearman's $\rho$ \\ \hline
        LVE & 0.166 \\
        PLRS & \textbf{0.437} \\\hline
    \end{tabular}}
\caption{\textbf{Human study on perceptual metric.} We conduct a human study to validate our proposed perceptual metric, PLRS.
We compute the Spearman's correlation coefficient $\rho$ to compare the PLRS rankings with the human preference rankings.
}
\label{tab:PLRS}
\end{table}

\section{Implementation Details of Ablation Study}\label{sec:F}
In this section, we provide implementation details of model variants ablated from our speech-mesh representation: the 3D SyncNet and the representation w/o 2D prior.

\paragraph{3D SyncNet}\label{para:3d_syncnet}
Inspired by Chung~\etal~\cite{syncnet}, we train 3D SyncNet to evaluate the performance of our transformer-based model compared to a CNN-based model.
3D SyncNet is trained using InfoNCE loss with a batch size of 80.
The architecture of 3D SyncNet consists of the mesh encoder comprising three dilated convolutional layers and the speech encoder with six convolution layers followed by two linear layers.
The mesh and speech features are extracted from each encoder, respectively.
We train 3D SycnNet on an NVIDIA RTX 3090 GPU for 20 epochs using LRS3-3D.
Also, for imposing the perceptual loss to 3D talking head models with 3D SyncNet, we finetune the model with VOCASET~\cite{voca} train split for 5 epochs, as our speech-mesh representation model does.

\paragraph{Ours w/o 2D prior}\label{para:w/o_2d_prior}
We train speech-mesh representation 
without Stage 1 training
to evaluate the effectiveness of our learned 2D prior. 
We train the speech encoder and mesh encoder, both with the same architecture as Stage 2, and the other hyperparameters are the same as in Stage 2. 

\section{Additional Results}\label{sec:G}
In this section, we present quantitative results on human studies (refer to \Sref{sec:G.1}) and Upper Face Dynamics Deviation (FDD) evaluation (refer to \Sref{sec:G.2}), comparing samples generated by the base models~\cite{faceformer,codetalker,selftalk} with and without perceptual loss to demonstrate the effectiveness of our speech-mesh representation. Additionally, we provide the qualitative result of temporal synchronization for the base models~\cite{faceformer, codetalker} (refer to \Sref{sec:G.3}).
We also provide comparisons on the stability of perceptual loss and cosine similarity for ablated model variants (refer to \Sref{sec:G.4}). 

\subsection{Human study on applying perceptual loss}\label{sec:G.1}
We conduct a human study to evaluate the perceptual preference for our method with two configurations: (1) training and testing on VOCASET, and (2) training on the combined MEAD-3D and VOCASET and testing on MEAD-3D, 
as
mentioned in Sec.~6.1 of the main paper.

In the first configuration, we ask participants, evenly balanced group of males and females with non-expert backgrounds, to compare two videos: one generated by the base model~\cite{faceformer, codetalker, selftalk} without our perceptual loss and the other with it.
To assess the quality of generated meshes, we design two separate descriptions—one focusing on lip synchronization and the other on overall quality.
For lip synchronization, participants are provided with the following description:  
``Please evaluate the lip synchronization between the speech and the lip movements in videos A and B, and choose the one that is more realistic and preferred.''  
A total of 18 participants take part in this evaluation. 
\Tref{tab:human_lr} shows that the participants significantly favor the models incorporating our perceptual loss with an overall preference rate of 72.9\%.
For overall quality, the description is as follows:  
``Please evaluate the overall quality of facial movements in videos A and B, and choose the one that is more realistic and preferred.''  
This evaluation involves 15 participants.
As shown in~\Tref{tab:human_overall_quality}, the participants show a strong preference for the model incorporating perceptual loss, with an overall preference rate of 73.3\%, indicating that the perceptual loss not only improves lip synchronization but also enhances the overall quality of facial movements.

In the second configuration, we ask 14 participants, also an evenly balanced group of males and females with non-expert backgrounds, to compare three videos: one generated by the base model~\cite{faceformer, codetalker, selftalk} trained on VOCASET, another generated by the base model trained on both MEAD-3D and VOCASET without our perceptual loss, and the other generated by the base model trained on both MEAD-3D and VOCASET with our perceptual loss. The description is as follows: ``Please rate the lip synchronization between the speech and the lip movements in videos A through C, with 3 being the most realistic and preferred, and 1 being the least.''
As indicated in \Tref{tab:human_exp}-(a) and (b), the participants significantly prefer the models incorporating MEAD-3D and our perceptual loss each by in 76.9\% and 67.9\% overall.
Notably, incorporating both MEAD-3D dataset and the perceptual loss results in 84.6\% of participants favoring the model, as shown in \Tref{tab:human_exp}-(c), compared to the original models.

This preference on the two configurations highlights the effectiveness of our speech-mesh representation as a plug-in module in enhancing lip synchronization from the perspective of human perception.

\begin{table}[t]
\centering
    
    \begin{tabular}{l@{\quad\,\,}c@{\quad\,\,}c}
    \toprule
     Model& w/o Our rep. & w/ Our rep.  \\
    \cmidrule{1-3}
    FaceFormer & 13.7\%& \textbf{86.3}\%\\
    CodeTalker & 32.4\%& \textbf{67.6}\%\\
    SelfTalk & 35.3\%& \textbf{64.7}\%\\
    \cmidrule{1-3}
    Overall & 27.1\%& \textbf{72.9}\%\\
    \bottomrule
    \end{tabular}
        
\caption{\textbf{Human study results on lip synchronization in configuration 1.} We adopt A/B test and report the percentage (\%) of preferences for A (Ours) over B, assessing the generated meshes on lip sync. Participants significantly favor the models incorporating our perceptual loss by in overall 72.9\%.}
\label{tab:human_lr}
\end{table}

\begin{table}[t]
\centering

\begin{tabular}{l@{\quad\,\,}c@{\quad\,\,}c}
\toprule
 Model& w/o Our rep. & w/ Our rep.  \\
\cmidrule{1-3}
FaceFormer & 14.4\%& \textbf{85.6}\%\\
CodeTalker & 27.8\%& \textbf{72.2}\%\\
SelfTalk & 37.8\%& \textbf{62.2}\%\\
\cmidrule{1-3}
Overall & 26.7\%& \textbf{73.3}\%\\
\bottomrule
\end{tabular}
\caption{\textbf{Human study results on overall quality in configuration 1.} We adopt A/B test and report the percentage (\%) of preferences for A (Ours) over B, assessing the generated meshes on overall quality. Participants show a strong preference for the models applying our perceptual loss, with an overall preference rate of 73.3\%.}
\label{tab:human_overall_quality}
\end{table} 

\begin{table*}[t]
\centering
    \resizebox{1.0\linewidth}{!}{
    \begin{tabular}{lcccccc}
    \toprule
    \multirow{2}{*}{Model} & \multicolumn{2}{c}{(a)}& \multicolumn{2}{c}{(b)}& \multicolumn{2}{c}{(c)}  \\
    \cmidrule(r{2mm}){2-3}    \cmidrule(r{2mm}){4-5}    \cmidrule(r{2mm}){6-7}
    & Original & Original + MEAD-3D & Original + MEAD-3D & Original + MEAD-3D + Our rep.& Original & Original + MEAD-3D + Our rep.  \\
    \cmidrule{1-7}
    FaceFormer & 33.3\%& \textbf{66.7}\%& 32.1\%& \textbf{67.9}\%& 19.2\%& \textbf{80.8}\%\\
    CodeTalker & 17.9\%& \textbf{82.1}\%& 34.6\%& \textbf{65.4}\%& 19.0\%& \textbf{91.0}\%\\
    SelfTalk & 17.9\%& \textbf{82.1}\%& 29.5\%& \textbf{70.5}\%& 17.9\%& \textbf{82.1}\%\\
    \cmidrule{1-1}\cmidrule{2-3}\cmidrule{4-5}\cmidrule{6-7}
    Overall & 23.1\%& \textbf{76.9}\%& 32.1\%& \textbf{67.9}\%& 15.4\%& \textbf{84.6}\%\\
    \bottomrule
    \end{tabular}}
\caption{\textbf{Human study results on lip synchronization in configuration 2.} We report the percentage (\%) of preferences for A over B, assessing the generated meshes on lip sync. Overall 84.6\% of participants prefer the model with MEAD-3D and our perceptual loss.}
\vspace{-4mm}
\label{tab:human_exp}
\end{table*}

\begin{figure}[t]
    \centering
    \includegraphics[width=1.0\linewidth]{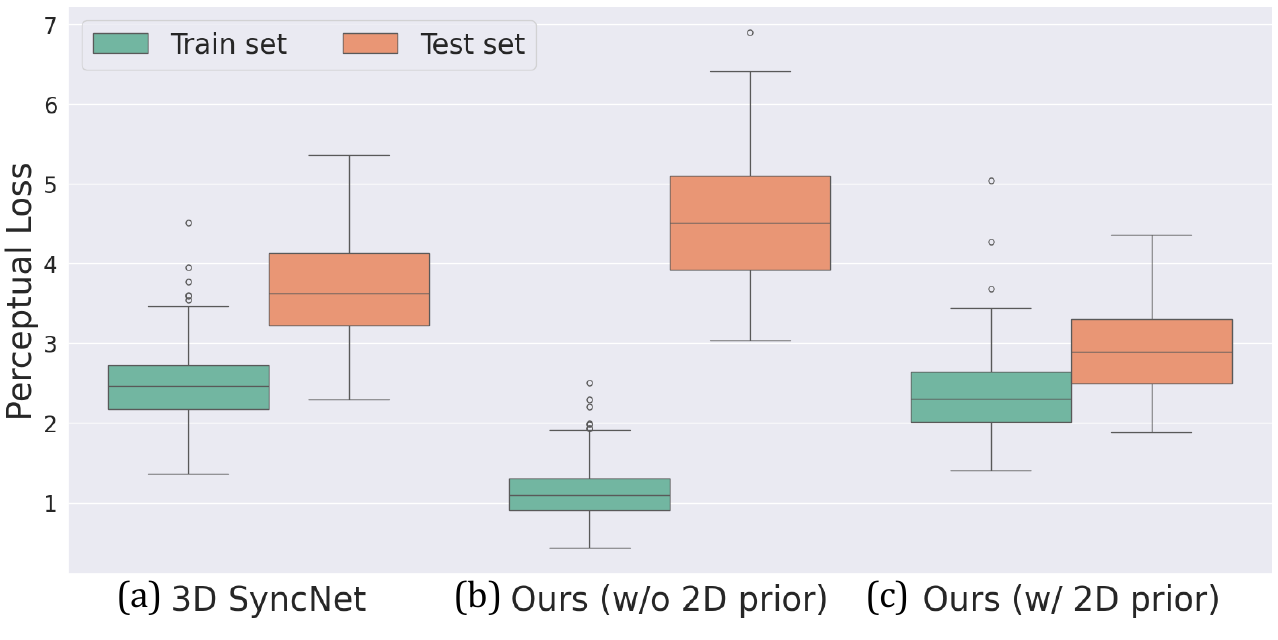}
    \caption{\textbf{Perceptual loss stability.} We visualize the perceptual loss between GT speech-mesh pairs on VOCASET samples. Our representation demonstrates strong generalization capability and provides a stable training signal compared to 3D SyncNet and our representation without 2D prior.
    }
    \label{fig:loss_stability}
\end{figure}

\begin{figure}[t]
    \centering
    \includegraphics[width=1.0\linewidth]{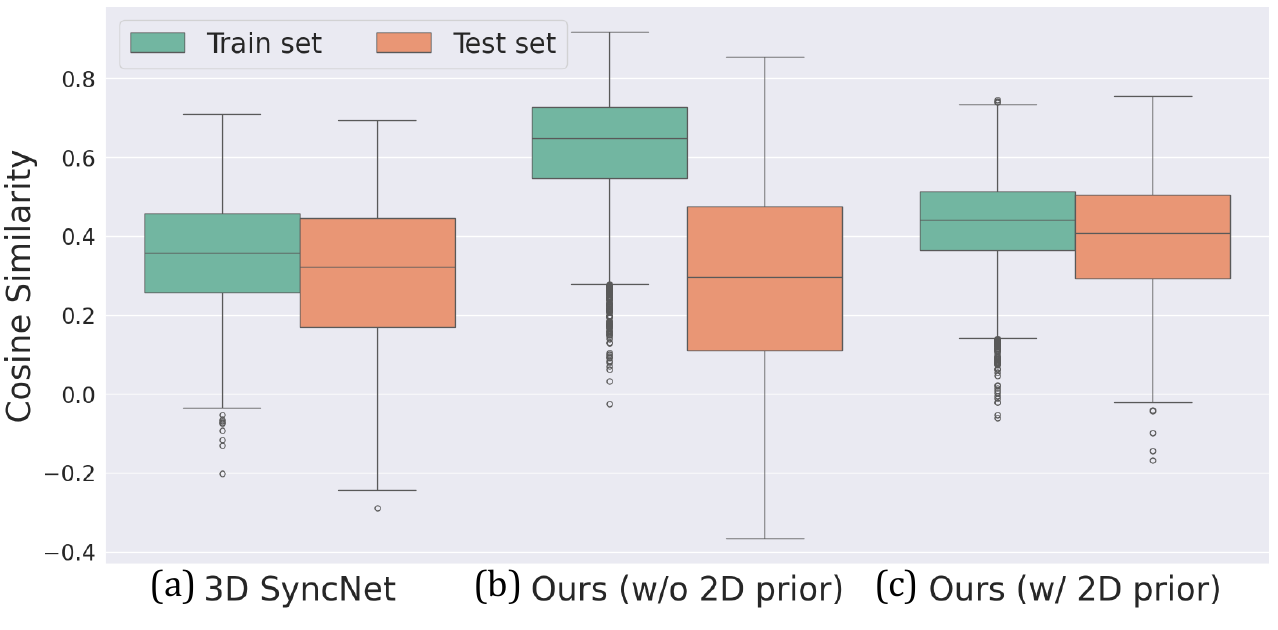}
    \caption{\textbf{Cosine similarity stability.} We visualize the cosine similarity between GT speech-mesh pairs on VOCASET samples. Our representation demonstrates strong generalization capability compared to 3D SyncNet and our representation without 2D prior.
    }
    \label{fig:cos_sim_stability}
\end{figure}

\subsection{FDD evaluation on applying perceptual loss}\label{sec:G.2}
In \Tref{tab:FDD}, we measure Upper Face Dynamics Deviation (FDD)~\cite{codetalker}, a widely used metric for the upper face evaluation, to assess the effectiveness of our perceptual loss. The models applying our perceptual loss achieve similar or improved FDD scores. It is expected because FDD is not the main focus of our work due to no direct relationship with the quality of lip movements.

\begin{table}[t]
\centering
\begin{tabular}{cc}
    \toprule
        & FDD $\downarrow$ \\
        & ($\times 10^{-7}\text{mm}$) \\
        \midrule
        FaceFormer & 3.789 \\
        + Ours rep. & \textbf{3.325}  \\\hline
        CodeTalker & 3.414  \\
        + Ours rep. & \textbf{3.259}  \\\hline
        SelfTalk & \textbf{3.319}  \\
        + Ours rep. & 3.424  \\
        \bottomrule
    \end{tabular}
    \caption{\textbf{FDD evaluation.} We report Upper Face Dynamics Deviation (FDD) scores to evaluate the variation in upper facial dynamics, which is not the main focus of our work. As expected, the models trained with our perceptual loss show similar or improved FDD scores.}
    \label{tab:FDD}
\end{table}

\subsection{Qualitative result of temporal synchronization}\label{sec:G.3}
\begin{figure}[t]
    \centering
    \includegraphics[width=1\linewidth]{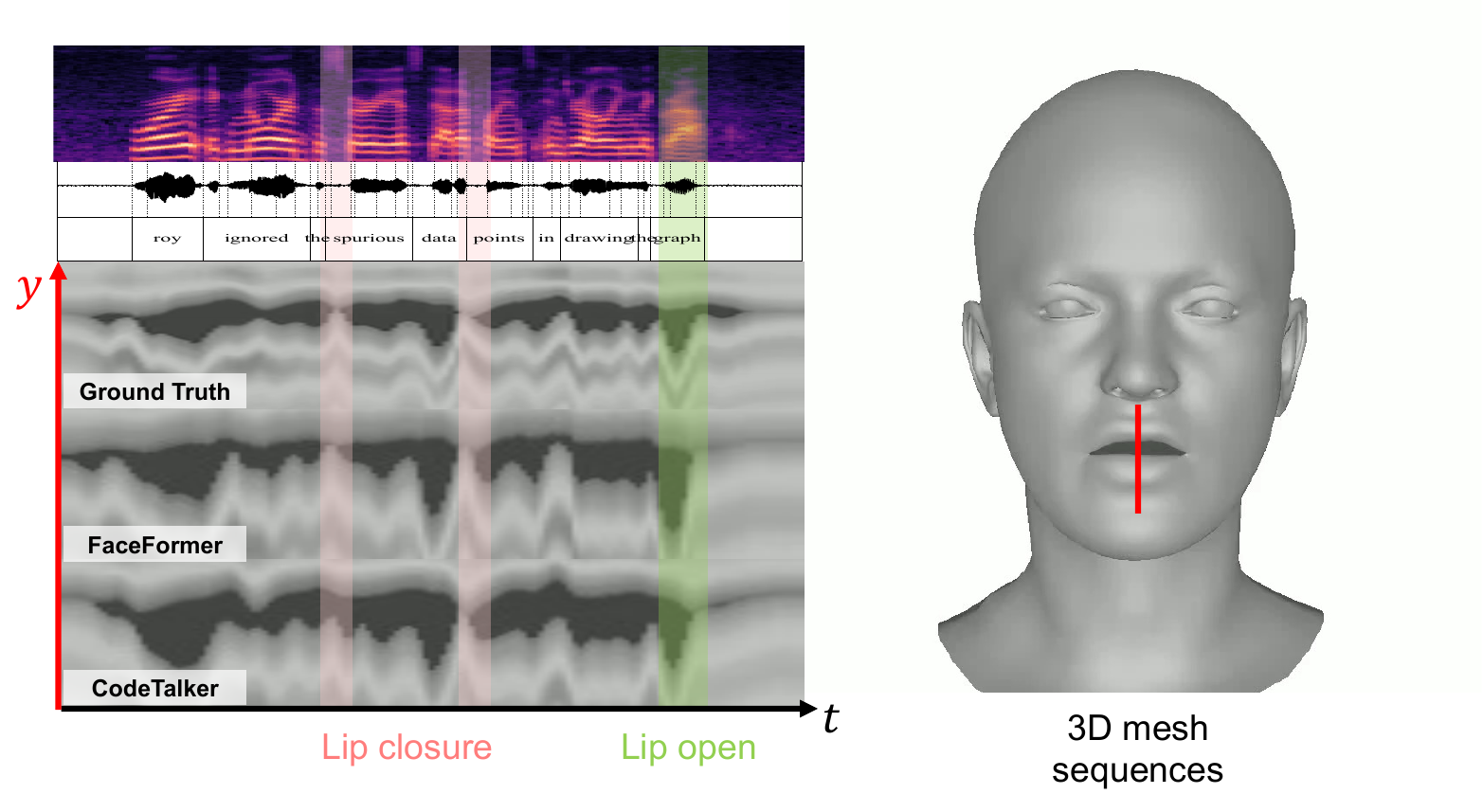}
    \caption{\textbf{Qualitative results of temporal synchronization on existing models. 
    }
    We plot y-t slices of rendered 3D face mesh sequences on the lip region with corresponding speech waveforms and mel-spectrogram.
    We also indicate the time steps of lip closure and opening with vertical lines.
    This implies that existing models already exhibit reasonable temporal sync.~capability.
    }
    \label{fig:temporal_qual}
\end{figure}

We present the qualitative result of temporal synchronization using existing base models~\cite{faceformer, codetalker, selftalk} (See Fig.~\ref{fig:temporal_qual}).
Given rendered 3D face mesh sequences, we place a vertical line with two pixel points near the lip region and extract the y-t slices of the mesh sequences to visualize the timing of lip closure and opening.
Next, we align the y-t slices with their corresponding speech waveforms and mel-spectrograms along the time axis. 
We observe that these models already have a reasonable temporal synchronization capability. 
Specifically, the timing of lip closure (\eg, for the /p/ sound) in the y-t slices aligns with minimal amplitude in both the speech waveforms and mel-spectrogram, while the timing of lip opening (\eg, for the /r/ sound) in the y-t slices coincides with a large amplitude in both speech representations.

\subsection{Stability comparison on loss and cosine similarity}\label{sec:G.4}
To utilize our speech-mesh synchronized representation as a perceptual loss, it is essential to provide a stable training signal to the 3D talking head model. 
In the domain of 2D audio-visual speech representation, Yaman~\etal~\cite{yaman2024audio} reveal that the transformer-based architecture~\cite{avhubert} learns more robust representation and provides more stable guidance to talking head models compared to a CNN-based approach~\cite{syncnet}.
To explore whether these observations hold for 3D speech-mesh representations, we evaluate both the lip-sync loss and cosine similarity across 3D SyncNet, our representation without 2D prior and our final representation.
This analysis aims to validate the effectiveness of the transformer-based architecture and curriculum learning with a pre-trained 2D speech representation.

Specifically, we measure the perceptual loss and cosine similarity, computing the mean and standard deviation for both the train and test samples.
Figures~\ref{fig:loss_stability} and~\ref{fig:cos_sim_stability}
show the comparisons of
perceptual loss and cosine similarity comparison across the three representation variants.
We denote the train samples as green box plots and test samples as orange box plots, respectively.

Our speech-mesh representation (Figs.~\ref{fig:loss_stability}-(c) and~\ref{fig:cos_sim_stability}-(c)) demonstrates the highest stability, exhibiting the lowest standard deviations (the height of the box plots) on test set in both lip-sync loss and cosine similarity.
In contrast, the representation without 2D prior (Figs.~\ref{fig:loss_stability}-(b) and ~\ref{fig:cos_sim_stability}-(b)) reveals significant discrepancies between the train and test samples on both the lip-sync loss and cosine similarity, indicating poor generalization capability.
Additionally, it shows the highest standard deviations, which potentially cause unstable training.
Meanwhile, 3D SyncNet (Figs.~\ref{fig:loss_stability}-(a) and~\ref{fig:cos_sim_stability}-(a)) displays the worst mean values of perceptual loss and cosine similarity among the three.

\begin{table}[t]
\centering
    \begin{tabular}{ccc}
        \toprule
        & Time $\downarrow$ & Mem. $\downarrow$ \\
        & (sec.) & (MB) \\
        \midrule
        FaceFormer & 0.447 & 1461 \\
        + Ours rep. & 0.537 & 1738  \\\hline
        CodeTalker & 0.138 & 3393  \\
        + Ours rep. & 0.289 & 3675  \\\hline
        SelfTalk & 0.175 & 8204  \\
        + Ours rep. & 0.320 & 8480  \\
        \bottomrule
    \end{tabular}
    \caption{\textbf{Training efficiency.} We compared the memory consumption and single-iteration speed during training with and without the perceptual loss. }
\label{tab:efficiency}
\end{table} 

\section{Discussion}\label{sec:H}
\paragraph{Limitations}
While our perceptual loss is applied only during training, which ensures that the resource requirements at inference remain unchanged, it requires additional computational resources during training. In \Tref{tab:efficiency}, we compare memory consumption and single-iteration speed during training, measured on a single A6000 GPU. 
Also, to capture the intricate correspondence between speech and 3D face mesh, we construct large-scale speech-mesh paired datasets, LRS3-3D and MEAD-3D. 
To this end, we utilize state-of-the-art monocular face reconstruction methods~\cite{spectre,retsinas20243d}, which may impose limitations on the quality of the 3D mesh in the reconstructed datasets.

\paragraph{Ethical considerations}
Our method can generate realistic 3D talking faces from arbitrary audio signals, relying on both the 3D scan data collected from actors and the reconstructed data from 2D talking videos.
Thus, while this technology has powerful applications, it also poses risks of misuse, such as creating harmful or embarrassing content.
To mitigate these risks, we emphasize raising public awareness and promoting ethical and responsible use through continued research.

\newpage

\begin{algorithm}[H]
\small
\caption{Mean Temporal Misalignment Calculation}
\label{alg:temporal_misalignment}
\begin{algorithmic}[1]
\footnotesize
\Require GT vertex sequence $V(t)$, Predicted vertex sequence $\hat{V}(t)$
\Ensure Overall mean temporal misalignment $\overline{\Delta t}$
\Statex
\State Initialize list of sample mean misalignments: $\{\Delta t_k\} \gets \emptyset$
\For{each sample $k$}
    \State Initialize time differences list: $\{\delta t_n\} \gets \emptyset$
    \State \textbf{Extract} lip vertices:
    \State \quad Upper lip vertex $V_u(t) \in \mathbb{R}^3$ from $V(t)$
    \State \quad Lower lip vertex $V_l(t) \in \mathbb{R}^3$ from $V(t)$
    \State \quad Predicted upper lip vertex $\hat{V}_u(t) \in \mathbb{R}^3$ from $\hat{V}(t)$
    \State \quad Predicted lower lip vertex $\hat{V}_l(t) \in \mathbb{R}^3$ from $\hat{V}(t)$
    \State \textbf{Compute} lip distance sequences:
    \State \quad $d_v(t) = \left\| V_u(t) - V_l(t) \right\|$
    \State \quad $\hat{d}_v(t) = \left\| \hat{V}_u(t) - \hat{V}_l(t) \right\|$
    \State \textbf{Smooth} sequences using \textit{Gaussian filter}:
    \State \quad $\tilde{d}_v(t) = \textit{Gauss}(d_v(t))$
    \State \quad $\tilde{\hat{d}}_v(t) = \textit{Gauss}(\hat{d}_v(t))$
    \State \textbf{Compute} derivatives:
    \State \quad $\delta \tilde{d}_v(t) = \tilde{d}_v(t) - \tilde{d}_v(t-1)$
    \State \quad $\delta \tilde{\hat{d}}_v(t) = \tilde{\hat{d}}_v(t) - \tilde{\hat{d}}_v(t-1)$
    \State \textbf{Perform} \textit{DDTW} to find alignment path $\mathcal{A} = \{(i, j)\}$
    \State \textbf{Identify} local extrema in $\tilde{d}_v(t)$ and $\tilde{\hat{d}}_v(t)$
    \For{each aligned pair $(i, j) \in \mathcal{A}$}
        \If{$i$ and $j$ are matching extrema of same type}
            \If{$j$ is within neighboring extrema range of $i$ in $\tilde{d}_v(t)$}
                \State Compute time difference: $\delta t_n \gets |i - j|$
                \State Append $\delta t_n$ to $\{\delta t_n\}$
            \EndIf
        \EndIf
    \EndFor
    \If{$\{\delta t_n\} \neq \emptyset$}
        \State Compute mean delta time for clip $k$:
        \State \quad $\Delta t_k = \dfrac{1}{N} \sum_{n=1}^{N} \delta t_n$
        \State Append $\Delta t_k$ to $\{\Delta t_k\}$
    \EndIf
\EndFor
\If{$\{\Delta t_k\} \neq \emptyset$}
    \State Compute overall mean temporal misalignment:
    \State \quad $\overline{\Delta t} = \dfrac{1}{K} \sum_{k=1}^{K} \Delta t_k$
\Else
    \State $\overline{\Delta t}$ is undefined
\EndIf
\end{algorithmic}
\label{algorithm:temporal}
\end{algorithm}

\end{document}